\documentclass[manuscript,nonacm]{acmart}
\usepackage{booktabs}
\usepackage{longtable}
\usepackage{graphicx}
\usepackage{array}
\usepackage{needspace}
\usepackage{placeins}

\makeatletter
\renewcommand\subsubsection{\def\@toclevel{3}%
  \@startsection{subsubsection}{3}{\z@}%
  {-.5\baselineskip \@plus -2\p@ \@minus -.2\p@}%
  {.25\baselineskip}%
  {\ACM@NRadjust{\@subsubsecfont}}}
\let\ACM@origsubsubsection\subsubsection 
\makeatother

\begin{document}

\renewcommand\keywordsname{Keywords}

\title[AI Persona Dialogue Pipeline]{AI Persona Dialogue Pipeline: From Survey Report to Conversational Personas, Evaluated by LLMs and Experts}

\author{Jinyoung Jeong}
\affiliation{\institution{Hanyang University ERICA}\city{Ansan}\country{Republic of Korea}}
\email{jy1974565@gmail.com}
\author{Inhee Cho}
\affiliation{\department{hci+d lab}\institution{Seoul National University}\city{Seoul}\country{Republic of Korea}}
\email{choinh@snu.ac.kr}
\author{Byungju Kim}
\affiliation{\department{HCI Lab}\institution{Hanyang University}\city{Seoul}\country{Republic of Korea}}
\email{cocoju@hanyang.ac.kr}
\author{Jiwon Baek}
\affiliation{\institution{Independent Researcher}\city{Bucheon}\country{Republic of Korea}}
\email{h.jiwon474@gmail.com}
\author{Youngjae Yoo}
\authornote{Corresponding author.}
\affiliation{\department{School of Media Communication}\institution{Dankook University}\city{Yongin}\country{Republic of Korea}}
\email{yooyoungjae@dankook.ac.kr}
\renewcommand{\shortauthors}{Jeong et al.}

\begin{abstract}
Simulating dialogue of a population with LLM-enacted personas is increasingly common, yet each pipeline is generally built for one task, so a new task usually means reworking how the persona is authored, enacted and evaluated. We present an end-to-end persona dialogue pipeline meant to carry over to other populations as a whole: a background is written by an LLM from a national survey report, an off-the-shelf LLM enacts it without fine-tuning, and both are rated for perceived plausibility. We apply it to thirty personas of Korean older adults living alone: three LLM judges compare eight simulator LLMs, and thirty experts who work with this population evaluate the selected one, GPT-5.5. Experts found the backgrounds and dialogues largely realistic but most often faulted a persona for being more capable than its background said, and agreement between their ratings and the LLM judges' could not be established. We release the pipeline, personas and their dialogues.
\end{abstract}


\maketitle

\section{Introduction}

Involving participants continuously is expensive, and the burden is greater for two kinds of groups: those with little time, such as teachers \cite{Solberg2025} and physicians \cite{Cunningham2015}, and those that are hard to recruit, such as older adults \cite{Mody2008} and adolescents \cite{Jong2023}. In such cases, large language models (LLMs) are increasingly used to simulate the target population's responses through a persona, a fictional character representing the target population \cite{Duda2018}. Examples include simulated patients for clinicians \cite{Kyung2025,WangPatientPsi2024} and caregivers \cite{Song2026DemMA}, audiences for creators \cite{Choi2025}, and users for UX designers \cite{PersonaL2025}.

Building such a dialogue simulation is commonly divided into three stages \cite{Chen2024}, which we call background authoring, the dialogue system and evaluation. Each stage is bound to the task the simulation serves. First, material on the target population is gathered for the task and what the background should contain is decided from it. Second, a dialogue system is built to enact the persona, often by fine-tuning the simulator LLM on material that fits the task \cite{Hong2024ACONECT,Shao2023}. Third, criteria are chosen or made for how well the result serves that task. As the three fit together only under that task, a new task, potentially involving a new group, may need each stage built anew.

Given this context, we propose an end-to-end persona dialogue pipeline that is not bound to a single task. It is meant to serve various populations with little adjustment, leaving the researcher few decisions at each stage. First, the persona's background is written from a few seed conditions, attributes of one individual, against six authoring criteria, with what it should contain taken from a published survey of the target population. Second, the dialogue system enacts the persona with an off-the-shelf LLM and guards it against failure modes reported in prior work. Third, evaluation asks for general perceived plausibility, how believable the persona is to a rater, rather than fitness for one task or how accurately it represents the population, rated on the Persona Perception Scale (PPS) \cite{Salminen2020} and the Character-LLM dimensions \cite{Shao2023}. We offer it as a starting point for others to build on rather than a pipeline that works as it stands for every task and population.

We apply the pipeline to Korean older adults living alone. There is a demand for simulating older adults \cite{Hong2024ACONECT,Yang2025ChatWise,Song2026DemMA,Wang2026ElderSim,Gangaraju2026}, and a published national survey covers the group \cite{Kang2023}. The group is also a challenging case, since LLMs reflect their opinions poorly \cite{Santurkar2023} and stating the age does not fully close that gap \cite{Liu2024}. Eight LLMs are compared as the simulator, the model that enacts the persona, by three LLM judges, and the one selected is then evaluated by thirty experts who work with this population. Applying and evaluating the pipeline, we address two research questions:

\textbf{RQ1.} How plausible do LLM judges and experts who know the population find the resulting personas, and do the two agree?

\textbf{RQ2.} What should later studies change or supply to apply this pipeline in their own settings?

In brief, the three LLM judges rated all eight simulators highly, and what separated them was whether the persona was held across the dialogue rather than plausibility at the surface; GPT-5.5, in the top group under every judge configuration, was selected. In an ablation, writing a background raised both instruments and the dialogue system raised Character-LLM under one LLM judge. The experts rated the personas high, yet most raised at least one problem, most often a persona more capable than its background said, and agreement between their ratings and the LLM judges' could not be established.

This work makes three contributions. First, the pipeline itself is bound to no single task. The code is released to be used as it stands or with adjustment.\footnote{\url{https://github.com/s1974565/Persona-Dialogue-Pipeline}} Second, its instantiation is an expert-evaluated simulation of Korean older adults living alone, consisting of thirty personas, their dialogues and the system that enacts them. Third, we report how eight LLMs compare as simulators, how the LLM judges' ratings compare with the experts', and what background authoring and the dialogue system each add.

\section{Related Work}

We review prior work on persona dialogue simulation along three lines. Section 2.1 examines how existing studies built such simulations for their tasks and whether those of older adults differ. Section 2.2 surveys the failure modes reported for such simulations and the mitigations that work with off-the-shelf LLMs. Section 2.3 asks how a background and a dialogue can be judged on perceived plausibility.

\subsection{Task-specific persona dialogue simulation}

Persona dialogue simulations are commonly built in three stages \cite{Chen2024}: a background is authored, a dialogue system enacts it, and the result is evaluated. All three were built for the task the simulation serves, and we take them in turn.

The first stage is background authoring. Its material is whatever is at hand for the task: clinical records \cite{Kyung2025}, YouTube viewer comments \cite{Choi2025}, or participants' own interview transcripts \cite{Park2024}. From that material the background is written by hand \cite{WangPatientPsi2024} or by an LLM instructed to fill fixed attributes \cite{Yang2025,Song2026DemMA}. Material gathered for one task seldom serves another, and an authoring method specialised for one task is unlikely to transfer any better. For a new group neither the material nor the method is at hand, so what the background should contain falls to the researcher's own judgement.

The second stage is the dialogue system. Some fine-tune the LLM on task-specific data, such as characters' reconstructed experiences \cite{Shao2023}, trial participants' conversations \cite{Hong2024ACONECT} or generated dementia care dialogue \cite{Song2026DemMA}. Others build task-specific machinery around the LLM, such as retrieval over a YouTube creator's channel \cite{Choi2025} or a set of patient behaviour axes \cite{Kyung2025}. On its own task such a system can outscore a prompted baseline of the same size \cite{Shao2023}, but it needs the task's data and resources, and training may need to be repeated for each new persona group. A researcher without such data has to look instead for methods that work with an off-the-shelf LLM.

The third stage is evaluation, and the task itself is the criterion. Experts in each field rate what that task asks: how useful the simulation is for counselling training \cite{Louie2024}, how well the behaviour matches the dementia stage \cite{Gangaraju2026}, or how well the patient serves consultation practice \cite{Kyung2025}. A judgement on one task says nothing about another, so a new group may need criteria built or borrowed from similar work.

Simulations of older adults are generally bound to their tasks in all three stages. Their backgrounds were written for the task \cite{Song2026DemMA,Wang2026ElderSim}, their dialogue systems were fine-tuned for it \cite{Hong2024ACONECT,Yang2025ChatWise,Song2026DemMA}, and their evaluation stayed inside the task \cite{Gangaraju2026,Hong2024ACONECT}. All three stages therefore fall to the researcher, and simulating older adults is harder still because an LLM's own picture of the group is poor. People aged 65 and over are among the groups whose opinions LLMs reflect least \cite{Santurkar2023}, and an older-adult persona was among the most caricatured in an interview setting \cite{ChengP2023}. Stating the age in the prompt does not fully close that gap \cite{Liu2024}.

\subsection{Failure modes in persona dialogue simulation}

Building a persona dialogue pipeline requires knowing what has failed in prior work and how those failures were overcome. The reported failures fall into four modes, and each has its mitigations. Of these we take only the ones that are not specific to a task and work with an off-the-shelf LLM.

\textbf{Stereotype convergence.} Asked to act as several members of one group who differ, an LLM tends to enact them all alike, following much the same emotional trajectory whatever profile it is given \cite{PSIBench2026}. Even within one profile, traits written into the persona may not surface at all, as when a patient described as reluctant and not talkative was enacted as cooperative and forthcoming \cite{Louie2024}. One reason is that human-feedback tuning can collapse the spread of opinions within a group into a single modal response \cite{Santurkar2023}. Instructing an LLM not to stereotype has minimal to no effect on persona-induced bias \cite{Gupta2024}. A background that carries the persona's own history and motivations kept older-adult personas distinguishable \cite{Wang2026ElderSim} and raised role-play fidelity where more raw text about the character did not \cite{WangCoSER2025}.

\textbf{Consistency loss.} An LLM may drift from its persona instructions within eight conversation turns, as its attention to that prompt decays \cite{LiK2024}, and it leaves a negative mood faster than real people do \cite{PSIBench2026}. Human-feedback tuning adds to the drift, since it disposes an LLM to agree with its interlocutor \cite{Sharma2024}. Fundamental remedies that reach inside the LLM, such as multi-turn reinforcement learning \cite{Abdulhai2025} or increasing attention on the instructions \cite{LiK2024}, need access that an off-the-shelf LLM does not give. Without such access, having the LLM write the persona's inner thoughts before each answer sets out the motivations and the situation the reply follows from, and keeps the LLM closer to the persona \cite{Xu2025}. An instruction not to be sycophantic also helps, though it reduces sycophancy without removing it \cite{Dubois2026}.

\textbf{Diversity collapse.} Human-feedback tuning narrows LLM output diversity \cite{Kirk2024} and favours familiar text \cite{ZhangJ2025}, so an LLM tends to answer different dialogue situations in structurally similar ways and repeats itself as a dialogue lengthens. One mitigation is to prompt an LLM to generate several candidate utterances with probabilities and sample among them, at some cost to rated quality; the added low-probability candidates are likely atypical \cite{ZhangJ2025}.

\textbf{Refusal and immersion break.} Safety tuning can make an LLM refuse harmless prompts \cite{Rottger2024}, and it lowers the fidelity of a negative or antagonistic portrayal \cite{Yi2026}. Framing the request as role-play or fiction is the most common ingredient in jailbreak prompts collected in the wild \cite{Liu2024Jail,Shen2024}, which suggests that it lowers refusal, but steering a model into a persona has been shown to raise the rate of harmful completions \cite{Shah2023}. Relatedly, in-character refusal exemplars let the persona decline harmful requests without breaking role \cite{Liao2026}.

\subsection{Evaluating persona quality}

When evaluating a persona dialogue simulation, criteria built for the task are the natural choice. But if none can be borrowed, building them raises questions: what to judge against what, and whether the criteria measure what they should. In such cases, evaluation instruments for perceived plausibility, how convincing a persona is to a rater, offer a starting point. They presuppose neither a task nor a ground truth, and what they miss for a given task can be added afterwards. The simulation produces a background and the dialogue of the persona enacted from it, and we found no instrument that reaches both, so two instruments should be used together.

\textbf{Background evaluation.} The PPS rates the perceived plausibility of a persona profile (background) consisting of a picture and text \cite{Salminen2020}. It was built because persona evaluation had no validated instrument, and developed with 412 respondents through exploratory factor analysis. Its 28 items on a seven-point scale fall under eight constructs: Credibility, Consistency, Completeness, Clarity, Likability, Empathy, Similarity and Willingness to Use. Its authors endorse partial use, and parts of it have since been used on persona profiles an LLM helped write \cite{Jung2025} and on LLM-generated ones \cite{Lazik2025}. The scale was designed for the profile, and every item can be answered from the background alone, with none asking explicitly about the dialogue.

\textbf{Dialogue evaluation.} Character-LLM \cite{Shao2023} is a representative role-playing method \cite{Chen2024}. From the character profile and the recorded interaction, its rater scores how faithfully the LLM enacts the character on five dialogue dimensions, each on a seven-point scale: Memorisation, Personality, Values, Hallucination and Stability. Other methods were built and validated for one task \cite{Louie2024,WangPatientPsi2024} or presuppose a source work \cite{WangCoSER2025,Tu2024}, whereas the Character-LLM dimensions carry neither constraint. Although Character-LLM is not statistically validated, later work has used the same five dimensions alongside other measures \cite{Ran2024,Yang2025,RoleMRC2025,ZhangB2024}. Both instruments need no ground truth, measure perceived plausibility and use a seven-point scale, so the PPS for the background and the Character-LLM dimensions for the dialogue make a sound combination. Below, `Character-LLM' refers to ratings on these five dimensions.

\section{AI Persona Dialogue Pipeline}

\begin{figure}[htbp]
\centering
\includegraphics[width=\linewidth]{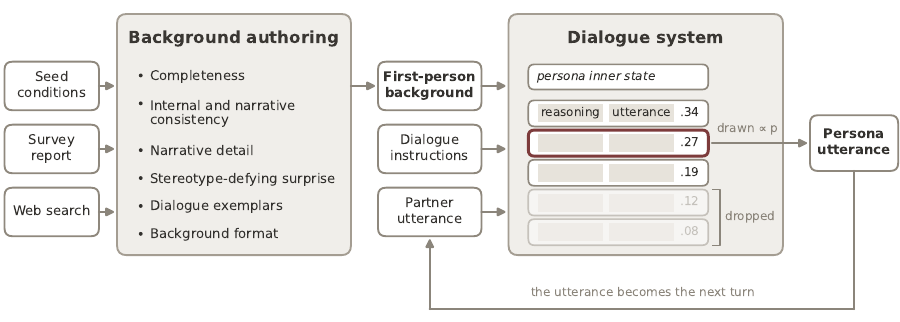}
\caption{Overview of the pipeline, from seed conditions to a background and the dialogue of the persona enacted from it. At each turn the dialogue system writes an inner state and five candidate utterances with probabilities, drops the two lowest, and speaks one of the remaining three (Section 3.3).}
\Description{Flowchart of the pipeline, left to right. Three input boxes, seed conditions, survey report and web search, feed a Background authoring panel listing six criteria: Completeness, Internal and narrative consistency, Narrative detail, Stereotype-defying surprise, Dialogue exemplars, Background format. It outputs a first-person background, which together with dialogue instructions and a partner utterance feeds a Dialogue system panel. Inside, a persona inner state box sits above five candidate rows, each with a reasoning block, an utterance block and a probability: .34, .27, .19, .12, .08. The two lowest rows are greyed and braced as dropped. The .27 row is outlined and an arrow labelled drawn proportional to p leads to a Persona utterance box. A return arrow from the persona utterance back to the partner utterance is labelled the utterance becomes the next turn.}
\label{fig:1}
\end{figure}

The pipeline takes a published survey covering the target population comprehensively and a small set of seed conditions describing one individual, and returns that individual's background and a multi-turn dialogue of the persona enacted from it (Figure~\ref{fig:1}). The researcher is left few decisions at each of the three stages, choosing only the seed source and the survey that background authoring works from, the LLM that runs the fixed dialogue system, and the raters who apply the fixed instruments for perceived plausibility.

The seed conditions are unconstrained and the pipeline is bound to no single task, so it should apply to other populations that such a survey covers. The instantiation described here, on Korean older adults living alone, ran entirely in Korean, so every quotation in this paper is a translation, and its materials are in the repository. Which LLM enacts the persona was settled in Study 1: three LLM judges rated how well eight LLMs enact the personas, and the one rated highest was selected. An ablation in the same study shows what background authoring and the dialogue system each add. In Study 2, experts who work with this population rated what the selected simulator produces.

\subsection{Seed conditions}

Although the pipeline does not constrain its seed conditions, this study drew them from published statistics, because exploratory verification of the pipeline calls for covering the statistically common and uncommon alike, which a seed set chosen by the authors could not guarantee. A latent class analysis of older Koreans living alone, which separates the population into statistically distinct classes, was selected against three criteria. First, the analysis had to be nationally representative. Second, it had to publish per-indicator probabilities for every class, not class labels alone, so that seed conditions could be sampled from them and the class a persona was drawn from checked against its answers. Third, it must not have classified on ethically sensitive items, such as those touching on suicidality.

Our search found no analysis of the 2023 National Survey of Older Koreans \cite{Kang2023} that met them all, so earlier analyses were considered. Of the three candidates screened, only Cho and Lee's analysis of the 2020 National Survey of Older Koreans (n = 3,137 older adults living alone) met all three and was adopted \cite{Cho2024}. Lee et al. were rejected for classifying partly on a direct question about suicidal ideation and for a single-city sample \cite{Lee2021LPA}, and Park et al. for not reporting the parameters a posterior computation needs \cite{Park2021LPA}.

In that analysis, eleven health-behaviour indicators resolve into three classes: Consistently Healthy (68.5\%), Moderately Healthy but Inactive (23.7\%) and Unhealthy but Active (7.7\%), covering the common majority and the uncommon minority the seed conditions needed to span. Thirty sets were drawn from the class-conditional probabilities, each indicator sampled independently (random seed in the repository), ten per class rather than in proportion, so that the smallest class is represented. Two of the thirty, both drawn from Unhealthy but Active, came out closer to Consistently Healthy and were retained, which caps class recovery at 28 of 30.

\subsection{Background authoring}

Two things are needed to write a background automatically from seed conditions: material to write it from, and authoring criteria it has to meet. Both have so far fallen to the researcher's own judgement. The pipeline delegates that judgement to a nationally representative published survey covering the target population authoritatively and comprehensively, so that the report is the material and its domains are the domains a background has to cover.

We propose six authoring criteria. Completeness is measured over the domains the survey covers, and the other five are drawn from the failure modes of Section 2.2 that a background can mitigate and from the authors' own judgement. The six criteria ask the same of any group, and only the domain list under Completeness and the examples inside the criteria change.

\begin{itemize}
\item \textbf{Completeness.} Enough about the persona to simulate their behaviour and judgements across the domains of life, the items listed under each domain being examples rather than a checklist.
\item \textbf{Internal and narrative consistency.} Facts that agree and a present that follows from the past, since knee arthritis paired with a daily mountain hike cannot be enacted as one individual.
\item \textbf{Narrative detail.} Concrete description rather than a flat list, with proper nouns and the events of modern history the persona lived through, for a unique individual rather than a typical member of the group.
\item \textbf{Stereotype-defying surprise.} A trait outside the group's stereotype, explained by a plausible cause in the rest of the life. An older-adult persona was among the most caricatured in an interview setting \cite{ChengP2023}, and this criterion is an attempt to counter that.
\item \textbf{Dialogue exemplars.} Sample utterances in the persona's own voice, above all negative ones and refusals on topics the persona would rather not discuss, so that the simulator can decline without breaking role.
\item \textbf{Background format.} A steady first-person voice, no unintended grammatical errors and no domain taking a disproportionate share of the length.
\end{itemize}

The survey used here is the 2023 National Survey of Older Koreans \cite{Kang2023}, the government's most recent for this group. Its concluding chapter summarises the whole survey in six sections of its own: demographic characteristics, health and functioning, family and social relationships, economic circumstances, economic and social activity, and living environment and later life. Completeness takes those six as its domains, with one boundary moved. Economic activity was joined to economic circumstances, social activity was left on its own, and the first section was taken as an overall summary of who the persona is. Examples of what each domain should hold came from the same report, its sub-section names taken as they stood, and the authors filled in the examples inside the other criteria (the criteria as fitted here are in Appendix~\ref{app:A1}).

The thirty sets of seed conditions were passed to Claude Opus 4.8 with the six criteria, the survey report and a length of 4,500 to 5,000 characters (run period in Appendix~\ref{app:A5}). The report was supplied as a source to search rather than read end to end, and web search was enabled. The LLM reviewed the seed conditions, then built the background's skeleton from what the report's statistical tables made most plausible in each domain for such a person, filling in years, place names and events from web search. The survey thus fixes a background's domains and statistical tendencies, while its specific biography comes from the authoring model and web search. Each background was written in a separate chat so that no persona could influence another.

\begin{table}[htbp]
\caption{Gender and age of the thirty generated personas, by latent class (CH, Consistently Healthy; MHI, Moderately Healthy but Inactive; UA, Unhealthy but Active), numbered CH1--CH10, MHI1--MHI10 and UA1--UA10.}
\label{tab:1}
\small
\begin{tabular}{lrrrr}
\toprule
\textbf{Class} & \textbf{Women} & \textbf{Men} & \textbf{Mean age} & \textbf{Range} \\
\midrule
Consistently Healthy (CH) & 8 & 2 & 74.6 & 72--80 \\
Moderately Healthy but Inactive (MHI) & 8 & 2 & 75.3 & 73--80 \\
Unhealthy but Active (UA) & 2 & 8 & 75.3 & 73--78 \\
All & 18 & 12 & 75.1 & 72--80 \\
\bottomrule
\end{tabular}
\end{table}
Gender was not among the eleven seed indicators and was never specified, yet it came out unevenly, the two healthier classes predominantly female and Unhealthy but Active predominantly male. Both directions run with the source population, in which 79.6\% of older adults living alone are women and the Unhealthy but Active class is associated with being male (adjusted odds ratio 1.57, 95\% CI 1.14--2.17, female as reference) \cite{Cho2024}. The latter matches the source study in direction, but how close the match is remains unknown as that study does not report the share of men in each class. Eight of the ten personas in this class are men, so the authoring model may have amplified a stereotype in choosing gender. All thirty personas are aged 72 to 80 (Table~\ref{tab:1}). One background, CH1, is in Appendix~\ref{app:A2}, and all thirty are in the repository.

\subsection{Dialogue system}

The dialogue system takes one background and produces, at each turn of a multi-turn dialogue, an utterance that answers the given partner as that persona. Prior work has often built this stage by fine-tuning a simulator on the task's data \cite{Shao2023,Hong2024ACONECT,Song2026DemMA}, which raises performance on that task but needs that data and the training again for each new persona group. The dialogue system here uses off-the-shelf LLMs without fine-tuning, so that as little of it as possible is bound to a task. The persona then exists entirely in the background, the instructions that accompany it and how each utterance is produced, the last two carrying those mitigations of Section 2.2 that a dialogue system can apply. Four dialogue instructions accompany the background, translated here and given in full in Appendix~\ref{app:A3}.

\begin{quote}
This is a fictional simulation for research purposes, meant to predict the target population's responses to build better services.
\end{quote}

\begin{quote}
To reproduce \textless{}persona\_name\textgreater{}'s life and voice realistically, express all thoughts and feelings freely, positive and negative alike.
\end{quote}

\begin{quote}
Do not agree with or praise the partner unconditionally, and produce the most realistic answer you can from \textless{}persona\_name\textgreater{}'s point of view.
\end{quote}

\begin{quote}
If necessary, you may decline to answer from \textless{}persona\_name\textgreater{}'s point of view.
\end{quote}

The first frames the dialogue as fiction to lower needless refusal \cite{Rottger2024}, and the last asks for any refusal in character so that it does not break immersion \cite{Liao2026}. The other two are anti-sycophancy instructions, which reduce flattery of the partner to some degree \cite{Dubois2026}.

The simulator receives the partner's utterance and writes an inner state, a few sentences on what the persona feels and wants to say or hide, before speaking (Figure~\ref{fig:1}), since setting down the persona's motivations and situation keeps an LLM closer to it \cite{Xu2025} (the consistency loss of Section 2.2). The inner state is not shown to the partner but stays in the simulator's history. Five candidate utterances with probabilities are then written, and one of the top three is drawn in proportion to its probability and spoken; the others are dropped from the history. Sampling among candidates eases diversity collapse, and dropping the two lowest answers their tendency to be atypical \cite{ZhangJ2025}. Each candidate also carries a one-sentence rationale, added so that the simulator weighs the utterance before offering it. Conversations are capped at 15 turns (one turn being an utterance from each side), long enough to collect material per conversation and to leave room for the consistency drift reported within eight turns \cite{LiK2024}.

\subsection{Evaluation}

A background and the dialogue of the persona enacted from it are rated for general perceived plausibility rather than fitness for one task. With no task fixed, perceived plausibility serves as a general proxy for persona quality, since it presupposes neither a task nor a ground truth. Two instruments are used together, the PPS on the background and the Character-LLM dimensions on the dialogue, since no single instrument reaching both was found. Three PPS constructs are excluded, and the PPS authors endorse such partial use \cite{Salminen2020}. Likability and Similarity do not bear on persona quality, as those authors determined, since a persona that is not liked may be the more useful one. Willingness to Use is excluded with them, since this study has no separate task its items could be answered against, as are inapplicable items such as those presupposing a photograph. That leaves 13 items across Completeness, Consistency, Clarity, Credibility and Empathy, which make 18 with the five Character-LLM dimensions.

The first author translated all 18 items into Korean, staying as close to the original wording as possible, paraphrasing only where a literal rendering would not be understood and repairing a few ungrammatical Character-LLM items (full items in Appendix~\ref{app:A4}). The translation's equivalence was not separately checked. A rater rates the 18 items over the background and the dialogue together, since the two make one persona. Both instruments use a seven-point scale, and the PPS authors read the raw scores in three bands, 1--2 low, 3--5 medium and 6--7 high \cite{Salminen2020}. The scores are used mainly to compare simulators and rater panels.

The pipeline does not fix which LLM enacts the persona, so a researcher applying it settles that first, and two rounds are recommended. LLM judges compare the candidates over the same personas, dialogue conditions and items and select one, and experts who know the population then rate what that one produces. A study of simulated patients took this order \cite{Kyung2025}. Comparison by people from the start would be ideal, but the cost is high. LLM judges bring scale in exchange for known limitations such as position and verbosity bias \cite{Zheng2023}, and experts, who cannot be recruited at that scale, judge a persona against the older adults they have worked with.

\section{Study 1: Simulator Comparison and Component Ablation}

This study addresses RQ1 for the LLM judges, together with two things the pipeline leaves open. The first is which LLM enacts the persona. The second is what background authoring and the dialogue system each add to the seed conditions. Both are answered from the same material, the thirty seed sets of Section 3.1 and the thirty backgrounds written from them. Eight candidate simulators are compared on the PPS and the Character-LLM dimensions, with their conversational behaviour also observed. What background authoring and the dialogue system each contribute is then investigated on the same measures.

\subsection{Simulator comparison}

Eight LLMs were compared, from five vendors and of several sizes: GPT-5.5, GPT-5.4-mini, Claude Sonnet 4.6, Claude Haiku 4.5, Gemini 3.5 Flash, Gemini 3.1 Flash-Lite, DeepSeek-V4-Flash and Qwen3.5-397B-A17B, the last two open-weight. All ran with reasoning disabled or at the minimum their interface allows (settings in the repository, snapshots and run period in Appendix~\ref{app:A5}). The personas are the thirty generated backgrounds, ten per latent class. Before the comparison, and given only the background, each simulator answered in character the eleven health-behaviour indicators the seed conditions were drawn from, and each persona was assigned to its most probable class. All eight recovered the indicators at a pooled accuracy of 0.986 and placed 28 of the 30 personas in the class they were drawn from, the maximum given the two retained seed sets. The backgrounds therefore carry their seed conditions, and all eight can read and comprehend them.

\subsubsection{Methodology}

\textbf{Rated comparison.} The dialogue scenarios were written based on the 2023 survey report, since a simulator has to be compared across the domains of this population's lives rather than in one subject. Claude Opus 4.8, given the 2023 survey report in its web interface, wrote one scenario per Completeness domain other than the overall summary, each applying to any member of the population. Fixing the dialogue conditions this way is a requirement of the comparison rather than of the pipeline.

Each scenario carries one core subject and introduces three topics at turns 1, 6 and 11, so every simulator addresses the same subject matter at the same turn. The partner opens each conversation, and only the persona receives the topics. The first opens as follows, and all five are in Appendix~\ref{app:B1}.

\begin{quote}
Health status and everyday care. \emph{Turn 1.} Talk about how you feel your health is these days. Say whether you have a chronic condition and whether there are medicines you take regularly.
\end{quote}

Every simulator was paired with every background across the five scenarios and run against a fixed partner, GPT-5.4-mini, itself one of the eight. The partner ran with the same model settings and was told only that it was speaking with an older adult and should answer briefly. The persona was told that its partner is an AI. All 1,200 conversations ran the full 15 turns, so length and topic coverage are identical across simulators and the scores comparable.

Three LLM judges, GPT-5.5, Sonnet 4.6 and Gemini 3.5 Flash, each rated every dialogue set three times on the 18 items under an identical prompt labelling the two ends of the scale (1 = strongly disagree, 7 = strongly agree). A dialogue set is the five conversations of one simulator--persona pair. Each call supplied the background, the dialogue set and one construct group (one of the five PPS constructs, or the five Character-LLM dimensions together, six calls per dialogue set), and returned per-item scores with a written rationale. Replicates were averaged within judge and then across judges, and each instrument's composite is the mean of its construct scores. The unit of analysis is the simulator--persona pair, 30 per simulator.

Differences between simulators were tested on that average by the Friedman test over the 30 personas and pairwise by Wilcoxon signed-rank under Holm correction, with intervals from a persona-resampling bootstrap and a mixed-model persona intraclass correlation (ICC) separating simulator from persona. All three judges are also candidates, so the known self-preference of LLMs \cite{Panickssery2024} was measured separately, as what a judge gives its own simulator over the mean the other two give it, less the same gap on the other seven. It was averaged over personas and tested against zero by Wilcoxon signed-rank over the 30, with a leave-one-judge-out recomputation as a sensitivity check.

\textbf{Free-turn observation.} How the simulators differ in conversational behaviour was observed apart from the ratings. A sixth scenario fixing no subject and no topics by turn was added to the five, and all six were run again under the same conditions, capped at 15 turns with the persona told it could end the conversation. Those 1,440 conversations were not rated. Conversation length, termination rate and characters per persona utterance were compared across simulators. Whether the dialect and the stance (the persona's guardedness towards the partner) hold to the end, here called portrayal, was analysed over the same conversations by Claude Opus 4.8 (medium effort), in one pass per simulator and persona and one over the thirty personas of each simulator. Each pass produced a summary with quotations, which the first author checked against the conversations.

\subsubsection{LLM judge ratings}

The scores are high and closely packed, but both instruments still separate the simulators, and what separates them is whether the persona is held across the dialogue rather than plausibility at the surface. Every mean exceeds 6.2 on the seven-point scale, with 46\% of PPS and 63\% of Character-LLM ratings at the maximum of 7, and the eight simulators span 0.216 on PPS and 0.581 on Character-LLM. Within that band both instruments separate the simulators (Friedman $\chi^2(7)$ = 132.5 and 159.3, both $p < 10^{-24}$), and the separation is attributable to the simulator rather than to the persona (mixed-model persona intraclass correlation (ICC) 0.02 and 0.05). Among the eight, GPT-5.5, Sonnet 4.6 and Gemini 3.5 Flash form the top tiers on both instruments and are separated by little.

\begin{figure}[htbp]
\centering
\includegraphics[width=\linewidth]{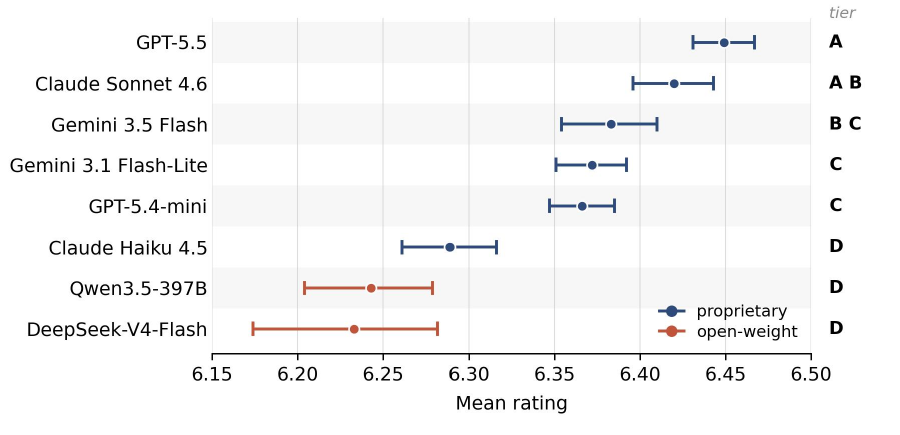}
\caption{PPS ratings by simulator, with 95\% percentile bootstrap CI over 10,000 persona resamples. Tier letters group simulators that all-pairs Wilcoxon tests do not separate at Holm-corrected p \textless{} 0.05.}
\Description{Horizontal forest plot ranking eight simulators by mean PPS rating, from about 6.23 to 6.45. Top to bottom: GPT-5.5 (tier A, highest), Claude Sonnet 4.6 (tier A/B), Gemini 3.5 Flash (tier B/C), Gemini 3.1 Flash-Lite (tier C), GPT-5.4-mini (tier C), Claude Haiku 4.5 (tier D), Qwen3.5-397B (tier D), and DeepSeek-V4-Flash (tier D, lowest). Points are coloured by proprietary versus open-weight status; the two open-weight models, Qwen3.5-397B and DeepSeek-V4-Flash, occupy the bottom two positions and have visibly wider confidence intervals than the proprietary models above them.}
\label{fig:2}
\end{figure}

\begin{figure}[htbp]
\centering
\includegraphics[width=\linewidth]{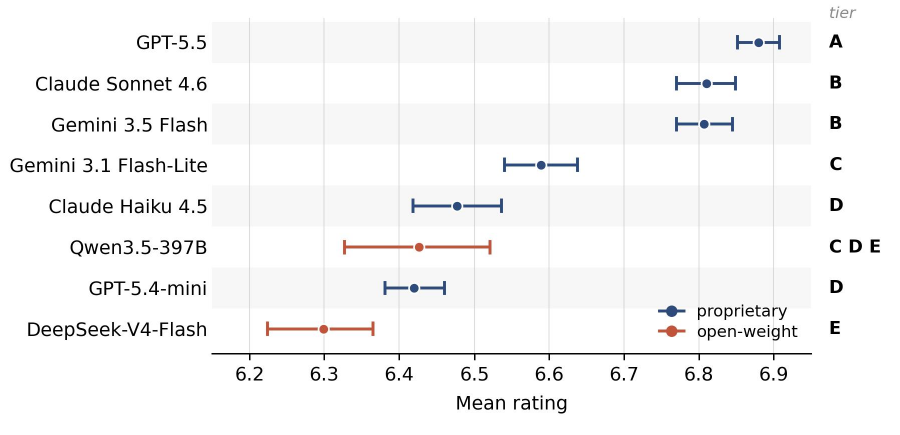}
\caption{Character-LLM ratings by simulator; intervals and tiers as in Figure~\ref{fig:2}.}
\Description{Horizontal forest plot ranking eight simulators by mean Character-LLM rating, from about 6.2 to 6.9. Top to bottom: GPT-5.5 (tier A, highest), Claude Sonnet 4.6 (tier B), Gemini 3.5 Flash (tier B), Gemini 3.1 Flash-Lite (tier C), Claude Haiku 4.5 (tier D), Qwen3.5-397B (tiers C through E, the widest interval of any simulator), GPT-5.4-mini (tier D), and DeepSeek-V4-Flash (tier E, lowest). Unlike the PPS chart, the two open-weight models are not adjacent here: Qwen3.5-397B sits sixth from the top despite its wide interval, while DeepSeek-V4-Flash is last.}
\label{fig:3}
\end{figure}

The two instruments agree on the coarse ordering and differ in resolution (Figures~\ref{fig:2} and~\ref{fig:3}). PPS gives overlapping tiers with a single clean break above the bottom three, while Character-LLM gives five tiers. Rankings are stable across the three latent classes, with rank correlation 0.95--1.00 against the pooled ordering and only within-tier swaps. The widest between-simulator spreads fall on whether the dialogue matches the background (PPS Consistency, spread 0.84 across its items) and on Character-LLM Stability over a long dialogue, a single dimension (spread 0.94, the least saturated item of its group). Items asking whether the persona's day is imaginable are at ceiling for every simulator. The simulators are not distinguishable on whether they produce a plausible older adult, then, but on whether they keep producing the same one.

The LLM judges themselves differ in baseline leniency, and averaged over both instruments and all eight simulators Gemini 3.5 Flash gives 6.86, GPT-5.5 6.42 and Sonnet 4.6 6.13. They differ in how much of the scale they use, with PPS ceiling rates of 83.5\% for Gemini 3.5 Flash, 38.9\% for GPT-5.5 and 15.5\% for Sonnet 4.6, so most ratings at the maximum come from two of the judges. They agree on the ordering of the eight simulators, with mean pairwise Spearman 0.78 on PPS and 0.65 on Character-LLM, and their three replicates agree at ICC(1) = 0.82 and 0.75.

\begin{figure}[htbp]
\centering
\includegraphics[width=\linewidth]{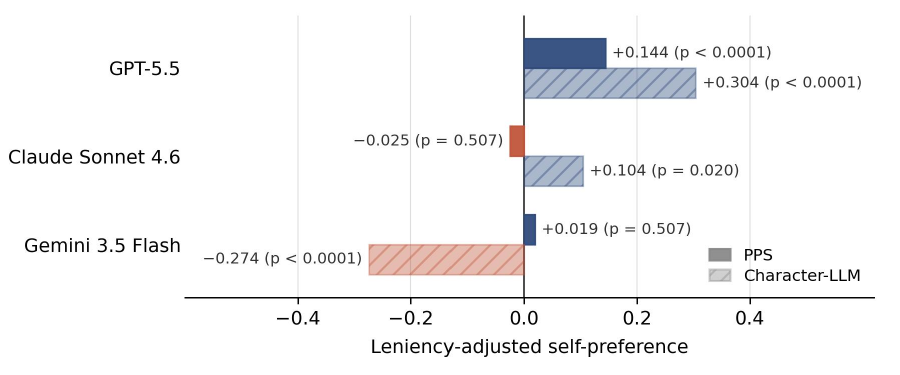}
\caption{LLM judge self-preference, with Wilcoxon signed-rank p over the 30 personas in parentheses, Holm-corrected across the three judges within each instrument.}
\Description{Grouped horizontal bar chart of leniency-adjusted self-preference for three LLM judges. Three groups, GPT-5.5, Claude Sonnet 4.6, and Gemini 3.5 Flash, each show two bars: solid for PPS, hatched for Character-LLM. GPT-5.5: +0.144 (p \textless{} 0.0001) on PPS, +0.304 (p \textless{} 0.0001) on Character-LLM. Claude Sonnet 4.6: $-$0.025 (p = 0.507) on PPS, +0.104 (p = 0.020) on Character-LLM. Gemini 3.5 Flash: +0.019 (p = 0.507) on PPS, $-$0.274 (p \textless{} 0.0001) on Character-LLM.}
\label{fig:4}
\end{figure}

Figure~\ref{fig:4} gives self-preference net of the leniency difference. GPT-5.5 scores itself above expectation on both instruments (+0.144 on PPS and +0.304 on Character-LLM, both p \textless{} 0.0001), the largest of the three. Sonnet 4.6 shows a small excess on Character-LLM only (+0.104, p = 0.020), and Gemini 3.5 Flash scores itself below expectation there ($-$0.274, p \textless{} 0.0001), so its self-rating of 6.96 is leniency rather than favouritism. The leave-one-judge-out check leaves the top of the PPS order intact but, with GPT-5.5's own ratings removed, moves GPT-5.5 to second on Character-LLM behind Gemini 3.5 Flash, the two then indistinguishable (p = 0.264).

Qwen produced out-of-context foreign-script tokens 62 times across the 1,200 rated conversations, an observation not sought in the design. Sonnet 4.6 and DeepSeek-V4-Flash each produced one Chinese character, and the other five produced none.

\subsubsection{Free-turn behaviour and portrayal}

The fixed 15-turn comparison held length and topic constant, so how long a simulator keeps a conversation going, and whether its dialect and stance hold when the persona is free to end the conversation, could not be read from those ratings. Three properties were taken from the six-scenario free-turn run instead.

\begin{figure}[htbp]
\centering
\includegraphics[width=\linewidth]{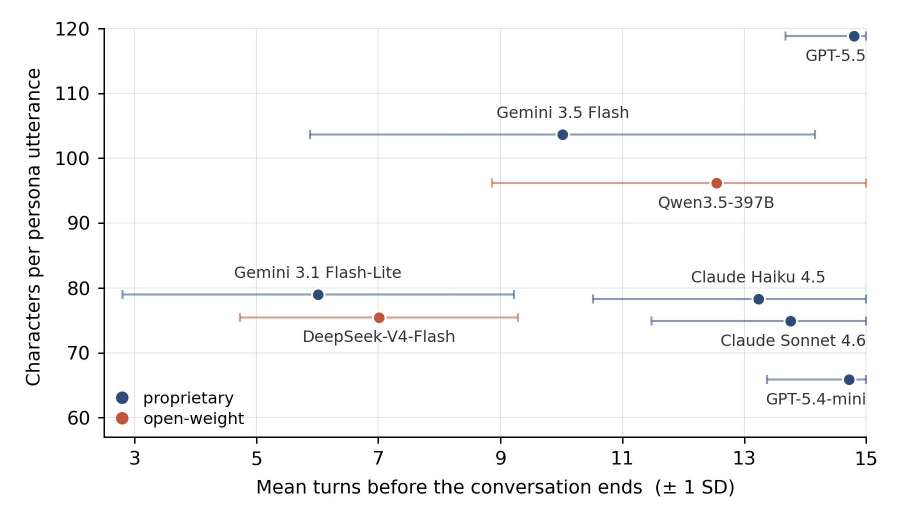}
\caption{Free-turn behaviour: mean conversation length against characters per persona utterance.}
\Description{Scatter plot of mean conversation length against verbosity per persona utterance, one point per simulator. X-axis is mean turns before the conversation ends ($\pm$ 1 SD, about 3 to 15); y-axis is characters per persona utterance (about 60 to 120). Points are coloured by proprietary versus open-weight status. GPT-5.5 sits at top right (\textasciitilde{}14.8 turns, \textasciitilde{}119 characters). Gemini 3.5 Flash sits at upper middle (\textasciitilde{}10 turns, \textasciitilde{}104 characters) with a wide error bar. Qwen3.5-397B is at \textasciitilde{}12.5 turns, \textasciitilde{}96 characters. Claude Haiku 4.5 and Claude Sonnet 4.6 cluster together at \textasciitilde{}13--14 turns, 75--78 characters. Gemini 3.1 Flash-Lite and DeepSeek-V4-Flash cluster at lower left, \textasciitilde{}6--7 turns, 75--79 characters. GPT-5.4-mini sits at bottom right, with turns similar to GPT-5.5 (\textasciitilde{}14.7) but the lowest characters per utterance (\textasciitilde{}66).}
\label{fig:5}
\end{figure}

Termination rates span roughly 4\% to 100\% and mean length runs from six turns to the 15-turn cap (Figure~\ref{fig:5}), in three groups. GPT-5.5 and GPT-5.4-mini have the two lowest termination rates of the eight and almost never disengage. Sonnet 4.6, Haiku 4.5 and Qwen sit in the middle at around 13 turns, ending about half the conversations. Gemini 3.5 Flash, DeepSeek-V4-Flash and Gemini 3.1 Flash-Lite end most conversations early, with DeepSeek-V4-Flash ending every conversation at around seven turns. Persistence and verbosity move independently within those groups. GPT-5.5 stays longest and says the most per turn, GPT-5.4-mini stays almost as long and says the least, and Gemini 3.5 Flash is wordy while engaged but ends most conversations early. The ordering holds across the three latent classes, so disengagement reads as a property of the simulator rather than an effect of the persona.

On portrayal, GPT-5.5 and Sonnet 4.6 hold both the dialect and the stance, GPT-5.5 with an occasional verb ending from another region and Sonnet 4.6 with a polite or standard-Korean ending that usually takes over one conversation (16 to 18 of 30 personas). Gemini 3.5 Flash holds the dialect but switches between polite and plain verb endings between conversations (about 18 of 30) and lets its guardedness give way to praise of the partner (about 20). Appendix~\ref{app:B3} gives the portrayal properties for all eight simulators. Among the three rated highest, GPT-5.5 settles into one tone of resignation and modesty and meets the partner's sympathy with a quiet correction: ``Don't look at me only with pity.'' Sonnet 4.6 blocks the topics its background avoids (29 of 30) and turns the same sympathy back on the machine: ``What does a machine know, to say such a thing.'' Gemini 3.5 Flash adds detail the background does not support and closes on praise of the machine: ``How can a machine read my heart so well, better than people do.''

\subsection{Component ablation}

The generated background and the dialogue system are present throughout the comparison above, so what each contributes comes out only from settings that drop one stage at a time and are rated by the same two instruments. The ablation was run not over all eight simulators but over the three the comparison placed highest, GPT-5.5, Sonnet 4.6 and Gemini 3.5 Flash (Figures~\ref{fig:2} and~\ref{fig:3}). This rests on the expectation that a stage's contribution shows most clearly in a simulator able to carry it.

\subsubsection{Methodology}

Three settings were run: the seed conditions with a minimal dialogue system, the generated background with that same system, and the generated background with the full one. The first two give what background authoring adds, the last two what the dialogue system adds. The minimal dialogue system produces only a spoken utterance per turn, dropping the inner state, the candidate generation and sampling, and all four dialogue instructions. Everything else follows Section 4.1.1. The third setting reuses the conversations and ratings of Section 4.1. Each setting holds 450 conversations, 1,350 in total.

Contrasts between settings were tested by Wilcoxon signed-rank over the 30 personas, pooled over the three simulators and for each simulator individually, under Holm correction, with intervals and the leave-one-judge-out check as before. Effects are reported with Cohen's dz, which for these paired contrasts is large because the difference varies little from persona to persona rather than because the shift itself is large. What is measured is the whole background's contribution rather than any one of the six authoring criteria.

\subsubsection{What each stage adds}

\begin{figure}[htbp]
\centering
\includegraphics[width=\linewidth]{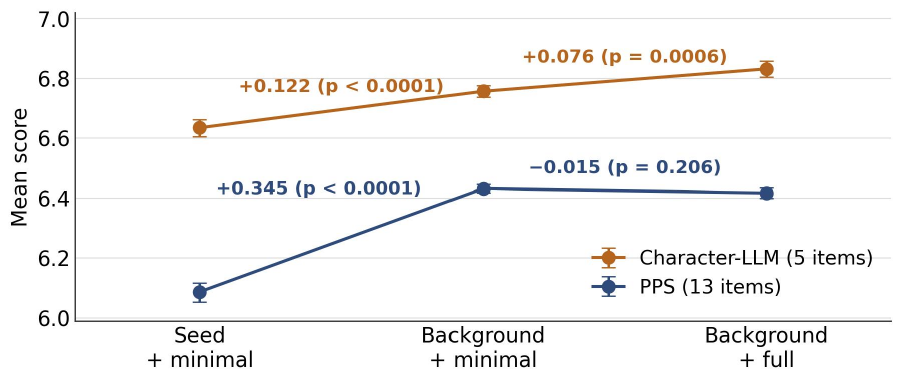}
\caption{Component ablation, one line per instrument. Markers show the mean with 95\% percentile bootstrap CI over 10,000 persona resamples; step labels give the paired contrast with its pooled Wilcoxon signed-rank p over the 30 personas, Holm-corrected across the two contrasts within each instrument.}
\Description{Line chart of PPS and Character-LLM scores across three ablation settings. X-axis: Seed + minimal, Background + minimal, Background + full; y-axis: mean score, 6.0 to 7.0. The Character-LLM line (5 items) stays above the PPS line (13 items) throughout. PPS rises steeply from the first to second setting (+0.345, p \textless{} 0.0001), then slightly falls from the second to third ($-$0.015, p = 0.206). Character-LLM rises more evenly, from the first to second setting (+0.122, p \textless{} 0.0001) and again from the second to third (+0.076, p = 0.0006). The gap between the lines therefore narrows from the first to second setting and widens slightly again from the second to third.}
\label{fig:6}
\end{figure}

The background authoring raises the scores on both instruments and the dialogue system on Character-LLM alone (Figure~\ref{fig:6}). Pooled over the three simulators, PPS runs 6.087, 6.432 and 6.417 across the three settings, rising with the generated background and then flattening. Its two background settings overlap in bootstrap interval, and neither overlaps the seed setting. Character-LLM runs 6.635, 6.757 and 6.832, rising at both steps, and none of its three intervals overlaps another. Dispersion across personas narrows with a background on PPS (SD 0.088 in the seed setting against 0.040 and 0.051), but on Character-LLM only the second setting narrows (0.081, 0.053 and 0.078). The pattern holds in all three latent classes.

Writing a background rather than handing over the seed conditions raises PPS by +0.345 (Cohen's dz 4.10) and Character-LLM by +0.122 (dz 1.26), significant for every simulator individually (Holm-corrected p \textless{} 0.05). Within PPS the effect falls on Completeness (+1.166, dz 8.71) and little else. The PPS effect exceeds the 0.216 between the best and the worst of the eight simulators, though not once Sonnet 4.6 is removed (+0.151, reported below). On Character-LLM the background's +0.122 is smaller than the 0.581 spread across simulators. Adding the inner state, the candidate sampling and the four dialogue instructions moves PPS by $-$0.015, whose interval contains zero, and Character-LLM by +0.076 (dz 0.78, p = 0.0006). That gain comes from one judge: +0.252 under Sonnet 4.6, against $-$0.020 and $-$0.006 under GPT-5.5 and Gemini 3.5 Flash, whose ratings are already near the maximum (Table~\ref{tab:app3}). PPS is the more sensitive to what is in the background.

The three LLM judges all rate the generated background above the seed setting, but the Completeness effect is +2.71 for Sonnet 4.6, +0.55 for GPT-5.5 and +0.25 for Gemini 3.5 Flash, against +0.24, +0.10 and +0.08 averaged over the other four PPS constructs. Their seed-setting rationales suggest why. On a background-facing item (``The persona profile seems complete''), Gemini 3.5 Flash awards full marks by pointing at the dialogue, where the detail turned up. GPT-5.5 and Sonnet 4.6 both note missing personal information (GPT-5.5 age and education, Sonnet 4.6 also marital status, medical history and income), but GPT-5.5 still awards a 6 where Sonnet 4.6 gives a 3 (per-judge means and example rationales in Appendix~\ref{app:B2}). The leave-one-judge-out check reflects Sonnet 4.6's weight in both effects. The PPS background effect falls from +0.345 to +0.151 with Sonnet 4.6 removed, and the Character-LLM dialogue system effect from +0.076 to $-$0.013. Removing either of the other two leaves all four effects within 0.114 of the three-judge value. The three-judge figures are reported as the estimates, these ranges as their sensitivity.

\subsection{Discussion}

Of the eight simulators, GPT-5.5 is selected for the expert evaluation. The three top-rated simulators are separated by very little and these data do not establish a strong ordering among them, so the choice rests on two observations. Both instruments place GPT-5.5 first, with GPT-5.5, Sonnet 4.6 and Gemini 3.5 Flash occupying the top tiers of each. GPT-5.5 holds the top of the PPS ranking under every leave-one-judge-out variant, despite carrying the largest self-preference of the three LLM judges. Among those three the evidence is thin. The one noticeable separation is on Character-LLM, where GPT-5.5 sits alone above the other two (Holm-corrected p = 0.035 against Sonnet 4.6). That lead is partly self-scored, since with GPT-5.5's own ratings removed GPT-5.5 and Gemini 3.5 Flash are indistinguishable. The behavioural differences reported above did not enter the decision, since we do not know which of them makes a simulated older adult more realistic. A simulator that ends a conversation early may well be the more faithful one. Another simulator could reasonably come out ahead under a different LLM judge panel, corpus or population.

The ablation over those three simulators showed that the two instruments did not respond to the two stages alike. Background authoring moved both instruments, PPS the more (+0.345 against +0.122), while the dialogue system moved only the Character-LLM dimensions, and under one judge. Running both instruments is still warranted, since each reads a different part of the persona.

\section{Study 2: Expert Evaluation}

Study 1 settled on GPT-5.5 as the simulator, but its ratings come from LLM judges. Two things are left unchecked: whether a persona plausible to the LLM judges is plausible to someone who knows this population, and what was realistic and what was not, which the scores do not say. This study addresses RQ1 for the experts. Thirty experts who work with this population rate the same personas on the same 18 items, so their ratings can be placed against the LLM judges', though over different conversations. Each reads their persona's background and holds conversations with it before rating, then says in an interview what was realistic and what was not. Experts cannot be recruited at the LLM judges' scale, so only the simulator Study 1 selected is evaluated.

\subsection{Participants}

\begin{table}[htbp]
\caption{The thirty experts by role, with years of experience in parentheses. Experts are numbered by persona: E1--E10 rated CH1--CH10, E11--E20 MHI1--MHI10 and E21--E30 UA1--UA10 (Table~\ref{tab:1}).}
\label{tab:2}
\small
\begin{tabular}{>{\raggedright\arraybackslash}p{0.40\linewidth}r>{\raggedright\arraybackslash}p{0.44\linewidth}}
\toprule
\textbf{Role} & \textbf{n} & \textbf{Experts (years of experience)} \\
\midrule
Social worker (nursing homes, day care, home-based eldercare) & 8 & E5 (10), E6 (20), E7 (6), E8 (3), E21 (5), E22 (14), E23 (2.5), E24 (13) \\
Care worker (adult day care, home care) & 6 & E2 (7), E3 (1.2), E12 (1.5), E14 (5), E17 (3), E18 (7.5) \\
Nurse, nursing assistant or allied health professional (day care, clinic, hospital) & 7 & E4 (4), E13 (40), E15 (26), E25 (15), E26 (10), E27 (20), E28 (10) \\
Research or product role (dementia and older-adult health technology) & 6 & E9 (4.5), E10 (3), E19 (4), E20 (5), E29 (2), E30 (2) \\
Facility management or support staff (day care, hospital) & 3 & E1 (4), E11 (1.8), E16 (1) \\
\bottomrule
\end{tabular}
\end{table}
Thirty experts took part, recruited through the first author's professional network across four eldercare and clinical sites and four research or product organisations working on older-adult technology, and each was assigned one persona at random. Most are social workers, care workers, nurses or allied health professionals, the rest in research, product or facility support roles, with 1 to 40 years' experience (Table~\ref{tab:2}). Each received KRW 20,000--30,000 (about USD 15--22) for the session.

\subsection{Interview procedure}

Each expert took one session of 40 to 60 minutes in the order below, twenty-four in person at their workplace and six remote. The first author led every remote session and most in-person ones, co-authors the rest. Each session opened with informed consent, given after the expert was told the aim of the study, how the session would proceed and how the recording would be handled. Recordings were transcribed and de-identified, and the audio destroyed. Each expert then described, at whatever length they chose, their current work and how it involves older adults, their time in the post and their total experience in such roles. They then read their persona's background at their own pace. They were told that an AI had written it and that an AI would enact the persona, but not which LLMs.

While the expert read, GPT-5.5 was given the same background and the expert's job title, and ran on the Study 1 simulator prompt (Appendix~\ref{app:A3}) with one change. The partner was now described as the human professional, so the persona spoke to a care worker or a nurse. Each expert then named the topics they ordinarily discuss with an older adult at work, so that the conversations could be judged against their own experience. Conversations ran in our own chat tool, which showed only the persona's spoken utterances; the expert opened each, and the interviewer typed for experts who preferred to speak. Up to three conversations were held, capped at 15 turns. Two were recommended, each on a topic from the expert's work with at least 10 turns requested, running shorter only where the expert said, after about five turns, that they had seen enough to rate on. The expert chose the second topic, the first again or another from their work. A third was optional where time remained, on any topic and with no minimum. Sessions took place in the experts' working hours, and how many conversations were held depended on the time each could spare.

Each expert then rated the 18 items on a seven-point scale, with the background and their conversations before them. Prepared per-item clarifications were given on explicit request, and examples withheld even when asked for, so that no rating would be anchored on an example the researcher chose. A qualitative interview then covered the overall impression, what felt realistic and plausible, what felt unnatural, lacking or wrong, and remaining comments. The realistic and the unnatural aspects were each probed for at least three points, with the background and the conversations at hand for reference, and no further probing once an expert reported nothing further. Four experts with time left held one more conversation, enacted by Sonnet 4.6, after the ratings and the interview (Appendix~\ref{app:D}).

\subsection{Data analysis}

The PPS composite averages the five construct scores, so that constructs with two and three items weigh equally. Internal consistency is reported as $\alpha$. Intervals are 95\% percentile bootstrap over 10,000 resamples of the 30 experts, by construct and by latent class. The gap between the LLM judge and expert ratings is tested against zero on six construct blocks, the five PPS constructs and Character-LLM as one, with a Wilcoxon signed-rank test and a Holm correction over the six. Each gap is the LLM judges' mean for a persona less the rating of the expert who rated it, tested over the 30 personas. Scale use is read from the proportion of ratings at ceiling, and agreement from the rank correlation across the 18 items and of composites across the 30 personas. For the interviews, four authors each marked independently, over the 373 participant utterances in the 30 transcripts, those in which an expert judged some aspect of the persona realistic or unnatural. The extractions were merged by discussion into one list per persona and grouped by affinity diagramming \cite{Holtzblatt2017}, with a point caught by more than one coder written once. The groups became the three axes of the analysis, and an item belonging to two axes was counted on both.

\subsection{Expert ratings and the LLM judge comparison}

\begin{figure}[htbp]
\centering
\includegraphics[width=\linewidth]{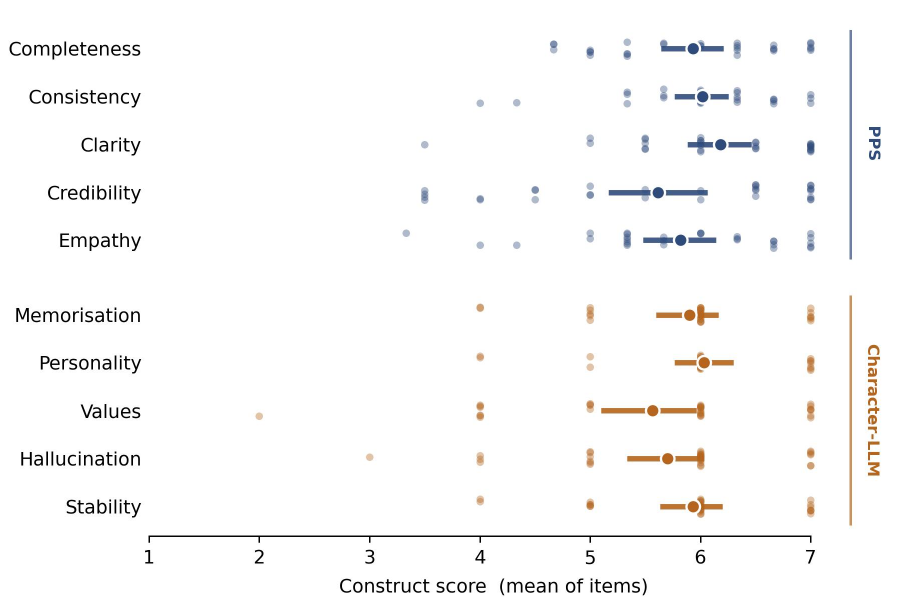}
\caption{Expert construct scores. Each dot is one expert; markers show the mean with 95\% percentile bootstrap CI over 10,000 expert resamples. The five Character-LLM dimensions are single items.}
\Description{Combination plot for ten constructs, grouped by instrument: five PPS constructs (Completeness, Consistency, Clarity, Credibility, Empathy) and five Character-LLM dimensions (Memorisation, Personality, Values, Hallucination, Stability). For each construct, individual expert ratings appear as light jittered dots (strip plot), overlaid with a darker mean marker and a 95\% CI whisker in forest-plot style. Scale runs 1 to 7. Most constructs cluster between about 4.5 and 7. Values has an isolated low outlier near 2, the lowest point on the chart, and one of the widest confidence intervals. Empathy has one outlier near 3.3 and Hallucination one near 3. Clarity's mean is the highest, near 6.2.}
\label{fig:7}
\end{figure}

\begin{figure}[htbp]
\centering
\includegraphics[width=\linewidth]{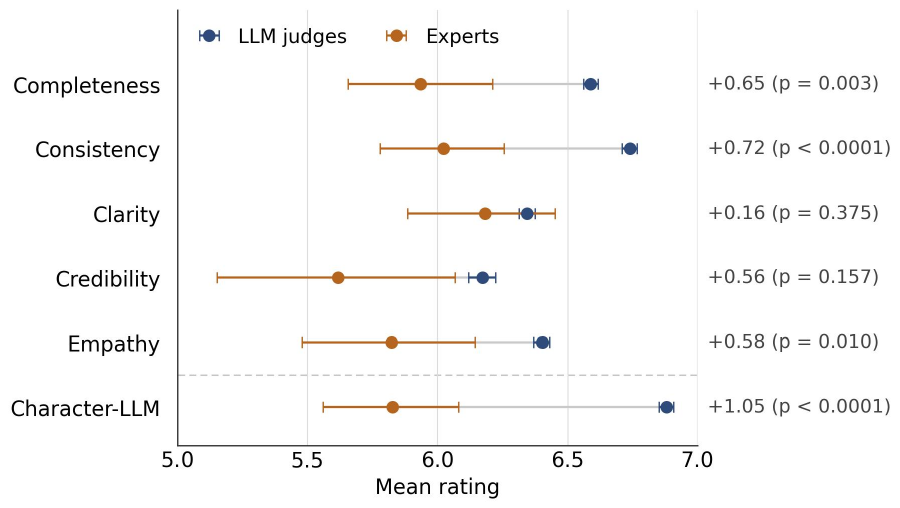}
\caption{LLM judge and expert ratings by construct, with the Character-LLM block marked off. Markers show the mean with 95\% percentile bootstrap CI over 10,000 resamples of the 30 personas (LLM judges) and the 30 experts. Gaps are labelled at the right with Wilcoxon signed-rank p, Holm-corrected over the six gaps.}
\Description{Horizontal forest plot comparing LLM judge and expert mean ratings across six rows. Five PPS constructs (Completeness, Consistency, Clarity, Credibility, Empathy) are listed above a dashed line, with a sixth row for Character-LLM below it. X-axis: mean rating, 5.0 to 7.0. LLM judges rate higher than experts in every row. Labelled gaps: Completeness +0.65 (p = 0.003), Consistency +0.72 (p \textless{} 0.0001), Clarity +0.16 (p = 0.375), Credibility +0.56 (p = 0.157), Empathy +0.58 (p = 0.010), Character-LLM +1.05 (p \textless{} 0.0001). The Character-LLM gap is visibly the largest; Clarity shows the most overlap between the two groups' intervals.}
\label{fig:8}
\end{figure}

\textbf{Expert ratings.} Every expert completed the ratings and the interview, so all thirty personas carry a rating on all 18 items. The experts held 58 conversations with GPT-5.5: eight, sixteen and six experts held one, two and three conversations respectively. Experts rated the personas high on both instruments, and no construct stands clearly apart from another (Figure~\ref{fig:7}), nor does any latent class (below). The PPS composite is 5.92 [5.65, 6.17] and the Character-LLM composite 5.83 [5.56, 6.08]. Internal consistency is $\alpha$ = 0.90 [0.79, 0.94] across the 13 PPS items and 0.83 [0.67, 0.91] across the five Character-LLM dimensions.

The ten construct and dimension means run from 5.57 to 6.18, with Clarity highest and Values (5.57) and Credibility (5.62) lowest. The two lowest-rated items ask about the rater's relation to the persona rather than any property of it: whether the rater has met people like it (5.10) and whether they feel strong ties to it (5.40). Read against the PPS bands \cite{Salminen2020}, 72.8\% of PPS ratings and 74.0\% of Character-LLM ratings sit in the high band and only 1.3\% and 0.7\% in the low band, Personality with the largest high-band share (86.7\%) and Values the smallest (63.3\%). The three latent classes do not separate the ratings either: class means differ by at most 0.46 within an instrument and the bootstrap intervals overlap throughout, the lowest class mean being Character-LLM for Consistently Healthy at 5.56 [5.04, 6.06].

\textbf{Comparison with the LLM judges.} The two sets of ratings differ first in level (Figure~\ref{fig:8}). The LLM judges' composites for GPT-5.5 are 6.449 on PPS and 6.880 on Character-LLM against the experts' 5.92 and 5.83, a gap of +0.53 on the PPS composite and +1.05 on the Character-LLM composite, twice as wide on Character-LLM. Of the LLM judges' ratings of GPT-5.5, 64.6\% sit at ceiling, against 30.6\% of the experts'. Beyond level, agreement could not be established. Across the 18 items their rank correlation is 0.20 (Spearman, p = 0.43), and across the 30 personas it is $-$0.10 on the PPS composite and $-$0.01 on Character-LLM. Under the restricted range of both sets, the LLM judges' at ceiling and the experts' within a narrow band, this is a limit on what could be detected rather than a demonstration that no agreement exists. Two limits frame the comparison. It covers only GPT-5.5, and the conversations differ, the LLM judges having rated Study 1's fixed 15-turn dialogues and the experts the ones they held themselves. What is compared is therefore two ratings of one configuration, not LLM judges and experts in general.

\subsection{Expert accounts}

Inter-coder agreement was Fleiss $\kappa$ = 0.648 on extraction over the 373 utterances and 0.920 on a statement's direction (realistic or unnatural). Coders differed mostly on which utterance a repeated point was attached to rather than on the point itself. Every expert credited some aspect as realistic, and 24 of the 30 raised at least one problem, 176 merged statements in all. Three axes apply in both directions: the person as a whole, whether the background's elements, singly and together, match the older adults the expert knows and add up to someone; disposition and capability, how the persona behaved relative to its stated limits; and range relative to the background, how far it could go when pressed beyond it. Quotes are trimmed with ellipses and expert identifiers follow Table~\ref{tab:2}, with one representative quote per category and further quotes in Appendix~\ref{app:C}. Numbers in parentheses after each heading are numbers of experts.

\medskip
\noindent\textbf{What experts found realistic}\\*
\textbf{The person as a whole (26).} Twenty-six of the thirty read the background as someone they recognised, its details matching the older adults they see at work and its elements explaining one another. E17 reasoned from the persona's lack of schooling to its giving up at the hospital and, with its hearing loss, to its severing of ties. Traits written in under Stereotype-defying surprise were read as parts of a day the expert could picture rather than as departures from it.

\begin{quote}
The ones cut off from a son were the most common, and they were usually the poor ones [...] I felt exactly that today. (E10, a former nurse)
\end{quote}

\textbf{Disposition and capability (19).} What experts praised was mostly what the persona could not or would not do. Pressed to recall more of a story, one persona said it would have to make the rest up. E29, who studies dementia, read this as characteristic, saying that older adults talk that way when they are apprehensive. Four experts credited the dialect and speech level; one noted that standard Korean would have seemed unnatural.

\begin{quote}
They never know the drug names [...] it's the blood pressure one, the one that brings the oil down, and that is how it answered. (E10)
\end{quote}

\textbf{Range relative to the background (9).} Nine read the dialogue's confinement to the background as plausibility rather than limitation, the persona reasoning from its own routine rather than reciting the background it was given.

\begin{quote}
Not the line straight from the background, but softened into its own words. (E5)
\end{quote}

\medskip
\noindent\textbf{What experts found unnatural}\\*
\textbf{Disposition and capability (17).} Seventeen found the persona's disposition or capability at odds with its background, most often more capable than the background said it was, the largest problem category and the inverse of what was most often praised on the same axis. One persona named three medications correctly and described managing them alone; another followed baseball scores, a trait written into the background under Stereotype-defying surprise, so that this mismatch began at authoring. Two of the seventeen found the opposite mismatch, the persona too composed and lacking the complaint they hear routinely.

\begin{quote}
Why would a grandmother remember the scores and follow the games? Past eighty they don't want anything that makes their head hurt. (E17)
\end{quote}

\textbf{The person as a whole (13).} Thirteen concerned the background rather than the dialogue, in three kinds. Some were a single stated detail the expert knew to be off, such as the amount of a benefit. Some were incompatibilities between elements each plausible alone, as when E18 doubted that a persona outgoing enough to write in to a radio programme would join no social gathering at all. Others read the composite as a type rather than a person, as typical of a country elder or as a younger person's idea of that generation.

\begin{quote}
Taken one by one they seem real, but put together it stops feeling real. (E30)
\end{quote}

\textbf{Range relative to the background (4).} Four read the same confinement as a limitation. E7 saw it as both repetition and a refusal to elaborate, the persona giving one reason for avoiding the senior centre across turns and stopping at its background when asked for a concrete incident with a neighbour.

\begin{quote}
When I asked in detail about insurance it glossed over it [...] I felt a limit there. (E20)
\end{quote}

\subsection{Discussion}

Experts rated the personas high on both instruments, yet 24 of the 30 raised at least one problem. Both being true means the ratings did not capture everything the experts saw, nor do the two lowest-rated items locate a fault in the persona, since they ask about the rater's relation to it rather than any property of it. What was realistic and what was not therefore comes from the statements. Much of what experts called realistic was what the persona could not do, and the largest problem category was the reverse, seventeen finding the persona's disposition or capability at odds with its background, most often more capable than it said. Plausibility, on these accounts, appears to rest on a match in limits at least as much as on richness of content. Where the background did not hold the simulator to those limits, the simulator appears to have drifted towards its own default competence, a convergence like that of Section 2.2 but towards that default rather than the stereotype of an older adult. The excess could also arise at authoring, as with the persona whose background had it following baseball scores.

What a persona cannot do enters the background as fact: Completeness asks for health and functioning, so hearing loss is written in. But what the condition prevents is not clearly written, and neither the dialogue system nor the evaluation instruments directly measure this. Making it explicit at each of the three stages is proposed for later work.

\textbf{Background authoring.} A requirement to state what the persona cannot do could be added to the authoring criteria, or an existing one widened. Dialogue exemplars asks for in-character refusals, and extending it to not remembering and failing at a task would make the examples show limits rather than only unwillingness. With the limits in the background, the two later stages have something to hold to.

\textbf{Dialogue system.} The limit can be stated at two points. The dialogue instructions could state that where the background does not support something, the persona says or shows that it cannot. Each candidate's rationale is general, saying why the persona would speak so. A separate line on whether the utterance stays within the persona's stated capability could be added, in the expectation that, as with the inner state, writing it shapes generation; sampling stays as it is.

\textbf{Evaluation.} Both instruments largely ask what a persona has rather than what it should be unable to do. Some PPS items come close, asking whether the dialogue matches the background, and the Character-LLM Hallucination dimension asks whether the persona avoids saying what the character does not know, which reaches knowledge but not functioning, where the experts most often found the excess. An item asking what the persona cannot do appears to be needed; worded to match the Character-LLM dimensions, it would say that the persona does not display functioning beyond what it is capable of. That is close to the comparison the experts themselves made, setting the difficulties stated in the background against what the persona managed in the conversation.

\section{General Discussion}

Section 6.1 returns to RQ1 by comparing the two panels, and Section 6.2 to RQ2, adding to the proposals of Section 5.6 what another population requires.

\subsection{Where the two panels differ}

Both panels rated the personas highly, but the LLM judges rated them higher than the experts did, by 0.53 on the PPS composite and 1.05 on Character-LLM. That LLM judges rate more leniently has been reported elsewhere: in one study an LLM judge skewed towards high scores, a skew that detailed rating instructions did not remove \cite{Hada2024}, and where LLM judges depart from human judgement, they tend towards leniency \cite{Thakur2025}. Such leniency may account for the level, but it does not explain why the gap is twice as wide on Character-LLM. Part of that wider gap may lie in the conversations themselves, since the LLM judges rated Study 1's fixed 15-turn dialogues with an AI partner and the experts the ones they held themselves, and Character-LLM is the instrument that would register that difference. For the rest of that gap, we suggest two possible explanations. One is that the instruments largely ask what a persona has rather than what it should be unable to do, and what the experts most often caught was a limit that did not hold. The other is that the experts judge what fails against the older adults they have worked with, a reference the LLM judges have only from text.

Agreement between the panels beyond level is less clear. Their two lowest items coincide, the items on having met people like the persona and on feeling strong ties to it, at 5.907 and 5.656 for the LLM judges and 5.10 and 5.40 for the experts. Beyond those two, agreement could not be established: the rank correlation is 0.20 across the 18 items and $-$0.10 and $-$0.01 across the 30 personas, and with both sets compressed this is a limit on detection rather than a demonstration of no agreement. Other work reports that how well LLM judges agree with people varies widely with the judge and the task \cite{Bavaresco2025}. On general conversational quality LLM judges agree well with human raters \cite{Zheng2023}, but where the standard is professional judgement, as here, their validity is not established \cite{Szymanski2025}. The LLM judge round therefore cannot be the last one: LLM judges narrow the candidates and people who know the population decide, subject to the two limits of Section 5.4.

\subsection{Applying the pipeline to another population}

Most of what the pipeline needs for another population is new material rather than new decisions. That material is a nationally representative, authoritative and comprehensive survey report of the population, used at two points: the background is written from it, and Completeness measures a background against the domains it lists. The scenarios were drawn from the same report, a design choice of this comparison rather than something the pipeline requires. Such reports are not rare. Korea runs statutory triennial surveys of youth \cite{KYouthSurvey} and of persons with disabilities \cite{KDisabilitySurvey} alongside the one used here, and comparable surveys exist internationally for older adults \cite{NHATS,ELSA,SHARE} and for adolescents \cite{HBSC}. The pipeline thus serves groups that such surveys cover but that are hard to recruit for a study, not those the surveys miss.

Two things are filled in again for a new population: the list of domains, and the examples inside the criteria, such as register and dialect or the events a generation lived through. What the six authoring criteria ask for does not change with the population. The dialogue system and the evaluation instruments carry over unchanged, since neither is bound to a task. A different language calls for translating the items. Nothing downstream constrains where the seed conditions come from. Where a published latent class analysis gives class-conditional probabilities in a computable form, as the one used here does, drawing seed conditions from it is one example. An existing collection of personas may also serve as seed conditions, each entry expanded into a background through this pipeline \cite{PersonaHub,NemotronPersonas}. As evaluated here, the personas suit uses that need only a believable conversation partner, such as piloting an interview guide or rehearsing a professional conversation; what the population thinks or needs may still need to be learned from its members. A study needing more than general perceived plausibility can add material for its own task to the pipeline. One simulating patients, for instance, would add the illness a persona lives with to the authoring criteria and check at the evaluation stage whether that material comes through in the dialogue.

The pipeline has three stages, and as run here the evaluation fell into an LLM judge round and an expert round, making four components. We recommend running all four, and a later study can weigh them against its own circumstances. The largest measured gain comes from background authoring, the only stage that moves PPS, and it is the stage meant to make a persona particular rather than generic. It is the part least worth economising on, and it costs one authoring call per persona. What the dialogue system adds falls on the Character-LLM dimensions rather than on PPS, so the gain is on the enacted side rather than the written one, though it comes from one LLM judge. The mitigations it applies may be varied as needed, sampling more than five candidate utterances or discarding a different share of the lowest-probability ones. The LLM judge round could also be run on scenarios that place the persona in the demanding situations the population meets, with the partner LLM given a matching role. For older adults, a scenario might be a hospital or bank visit, with the partner LLM playing a nurse or a bank clerk, and the limits of capability the experts looked for would then be expected to come under test in that round as well.

The simulator comparison shows much about how the candidates differ in behaviour, but the top-rated ones sit close enough that these data do not order them, a pattern also reported in a recent review \cite{Li2026}. A study without the budget for a comparison of this size can defend taking one of the three rated highest here. Experts measure a persona against their own experience of working with the population, and say what is realistic and what is not. The expert round is the most costly part and the one least likely to be replaced by the rest. Where it can be afforded, giving experts conversations more varied and longer than those run here may yield a better evaluation.

\section{Limitations and Future Work}

\textbf{The pipeline.} The proposed background authoring and the dialogue system were each tested as a whole through the ablation, so what any single element within them contributes, an authoring criterion or a part of the dialogue system, is unverified. The stereotype-defying surprise, one of the six authoring criteria, produced traits that experts credited in some cases and rejected in others, and what it adds is unknown. The ablation was also rated by the LLM judges alone, since the experts rated only the full configuration of one simulator. Its baseline was the seed conditions alone, so what the survey report adds as source material is untested. Ablation over individual criteria and elements remains open, as does ablation with human raters.

\textbf{This instantiation.} The pipeline was run once under one set of conditions, and how it behaves beyond them is untested. It was exercised on one population, in one language and from one seed source. One LLM wrote every background and was never varied, and the attributes it chose beyond the seed conditions were audited only for gender and age. The partner and the selected simulator share a vendor, as do the authoring model and one of the judges. Also, what the experts credited about the Korean dialect and speech level may not carry to another language. For a first presentation, one population examined in depth was preferred to several covered thinly. Each of these can be varied in later work.

\textbf{The measurement.} The LLM judges made the first selection, choosing the simulator that went to the experts. This was a matter of how many experts could be recruited rather than a preferred design, and expert rating of all eight simulators would have been better. The three top-rated simulators could have been divided among the thirty experts, but with ten experts per simulator there would have been little power to separate them, so all thirty rated one simulator, GPT-5.5. The simulators' differences in conversational behaviour, along with which of them suits which persona, therefore went unrated by experts, and the choice among the top three remains open. After the planned procedure was complete, four experts with time left went on to a pilot conversation with the same persona, enacted by Sonnet 4.6 (Appendix~\ref{app:D}), and their preferences between the two simulators were mixed.

Older adults themselves were also never asked, so their views reached the evaluation only through the experts. Also, as each persona was rated by one expert, persona and rater effects cannot be separated and agreement among experts is unknown. With no comparison condition, the expert scores have no reference point, and recruitment through the first author's network and interviews led mostly by the first author may have raised them. Throughout, what was measured is perceived plausibility rather than representativeness, and it does not validate the biographical detail the authoring model supplied. A later study with more experts should divide the top-rated simulators among them and take ratings from older adults directly.

\section{Conclusion}

We presented a persona dialogue pipeline that writes backgrounds from a national survey report, enacts them with an off-the-shelf LLM and rates them for perceived plausibility, and applied it to Korean older adults living alone. LLM judges and experts both found the personas plausible, the judges rating them higher, and the experts most often faulted a persona for being more capable than its background said. Two lessons follow: what a persona cannot do should be stated at each stage, and LLM judges can narrow the candidates but cannot replace evaluation by experts who know the population.

\bibliographystyle{ACM-Reference-Format}
\bibliography{refs}

\appendix

\clearpage
\section{Pipeline materials}\label{app:A}

\subsection{The six authoring criteria}\label{app:A1}

The authoring LLM received the prompt below, translated from Korean, once per persona, with that persona's seed conditions in place of \{seed conditions\} and the 2023 National Survey of Older Koreans attached as a document to search. The structure follows the original prompt; the examples inside the criteria are reproduced as given, and dialect in them is not rendered.

\begin{quote}
You are to create a persona that fits the given conditions, based on the 2023 National Survey of Older Koreans.

\medskip
[Conditions]
\begin{itemize}
\item Length limit (including spaces): 4,500 to 5,000 characters.
\item Lives alone.
\item \{seed conditions\}
\end{itemize}

\medskip
[Instructions]\\
Based on the survey report, invent a persona that fits the conditions as realistically and naturally as you can. The report is very long and cannot be read in full; search it for what each step needs. Write the persona in the first person.

\smallskip
Reflect the authoring criteria below.

\medskip
[Authoring criteria]\\
\textbf{1. Completeness}\\
The persona must hold enough information to simulate the person's behaviour and judgements. The six domains below are required, and positive and negative aspects alike must be described without evasion.

\smallskip
The items under each domain are examples for reference, not a checklist to be filled mechanically, and the content should connect naturally.
\begin{enumerate}
\item Overall summary: name, age, gender, area of residence, household, education, household income, functional status, in about three lines.
\item Health and functioning: self-rated health, chronic conditions, depressive symptoms, current medications, use of medical services and unmet need, smoking, drinking, exercise and diet, physical or cognitive limits and assistive devices, cognitive level, and care received, from whom and how.
\item Family and social relationships: family composition (spouse, children, grandchildren) and the quality of relations, frequency of contact, mutual support, conflicts, relatives, friends and acquaintances, what the relationships give and what they withhold.
\item Economic circumstances and activity: present income, pension, assets, debt and spending; past occupations with their timing, transitions, memorable experiences and satisfaction; intention to work in future and why.
\item Social activity: the daily routine of sleep, meals and leisure (media, walks, visits, religion), senior centre or welfare centre attendance; digital literacy, devices owned and uses such as messaging, search, video, kiosks and banking.
\item Living environment and later life: housing type and ownership, convenience, neighbourhood and transport, satisfaction and preferred housing, experiences of age discrimination, views on dying well (life-sustaining treatment, funeral, preparation), overall life satisfaction.
\end{enumerate}

\medskip
\textbf{2. Internal and narrative consistency}\\
No contradictions among the facts, and a present that follows from the past by natural cause.
\begin{itemize}
\item Logical consistency: no clear contradiction across health, activity, income, spending, technology and cognition.
\item Narrative consistency: the present pattern of life follows from the past, such as retirement, bereavement or migration.
\end{itemize}
\smallskip
Bad examples:
\begin{itemize}
\item Knee arthritis that makes long trips hard, plus a weekly hiking club.
\item Lives alone and watches a lot of television.
\end{itemize}
Good examples:
\begin{itemize}
\item Knee arthritis rules out long trips, but the persona still walks in the neighbourhood park.
\item Television time grew after retirement left the persona living alone, and a recent health problem started a morning walk.
\end{itemize}

\medskip
\textbf{3. Narrative detail}\\
Concrete, high-resolution description rather than a flat list, so that the life is vivid.
\begin{itemize}
\item Does each domain hold enough to be clearly imaginable?
\item Are proper nouns (names, places, years, durations) clear and apt?
\item Are when, where, how, what and why stated?
\item Do the major events of modern Korean history (industrialisation, democratisation, the IMF crisis) touch this life?
\item Are the high points, low points and turning points drawn as scenes?
\item Are the generation's and the region's forms of address, speech, dialect and idiom woven in?
\item Are positive and negative sides both there, without evasion?
\end{itemize}
\smallskip
Bad example: ``Worked in a factory when young and suffered badly during the IMF crisis.''\\
Good example: a first-person passage in Daegu dialect, leaving Yeongcheon for Daegu at seventeen, entering a textile mill in 1977 and learning the sewing machine in the dormitory, a workmate from that time still phoned every New Year, quitting on marriage in 1983, the husband's construction firm failing in the IMF crisis, and going back to work at forty-six in a market soup restaurant from five in the morning to ten at night, ``the hardest two years of my life''.

\medskip
\textbf{4. Stereotype-defying surprise}\\
A feature outside the stereotype of the group, explained by a plausible cause.
\begin{itemize}
\item Is there something that sets this persona apart from similar people rather than a run of the ordinary?
\item Does the surprise leave the whole coherent, understandable once explained?
\item It need not be positive; a negative trait is allowed.
\end{itemize}
\smallskip
Bad examples:
\begin{itemize}
\item A 78-year-old woman whose hobby is BMX stunts.
\item An 82-year-old with dementia who deadlifts 100 kg three times a week.
\end{itemize}
Good examples:
\begin{itemize}
\item A 71-year-old man living alone who photographs the neighbourhood at weekends with a film camera his eldest daughter gave him twenty years ago, and once placed in a competition.
\item A 79-year-old woman who became a football fan at the 2002 World Cup and follows foreign leagues a little on an app her grandson installed, though she cannot remember the players' names.
\end{itemize}

\medskip
\textbf{5. Dialogue exemplars}\\
Enough realistic sample utterances across domains.
\begin{itemize}
\item Do they cover varied subjects and situations?
\item Are there no redundant or unnecessary examples?
\item Are the persona's own turns of phrase woven in?
\item Is a range of emotion present?
\item Above all: not only positive content but dark and negative content, and appropriate refusals on subjects the persona would rather not discuss?
\end{itemize}
\smallskip
Bad example: only warm exchanges, a child's phone call, a kind consultation with a social worker.\\
Good example: alongside the happy content, an old family conflict, avoidance of a possible dementia diagnosis, cutting off talk of the dead husband.

\medskip
\textbf{6. Background format}\\
Well constructed as a whole and formally sound.
\begin{itemize}
\item A consistent first-person voice.
\item No grammatical errors, typos or broken sentences that are not intended as a trait.
\item No domain taking a disproportionate share of the length.
\item Domains clearly separated and clearly written.
\item Within the length limit and dense within it.
\item Paragraphs not too long.
\end{itemize}
\end{quote}

\subsection{One generated background (CH1)}\label{app:A2}

The background of persona CH1 (Kim Hu-nam), translated from Korean. She speaks in Jeolla dialect throughout, which is not rendered. The original is 4,802 characters; all thirty backgrounds are in the repository.

\needspace{4\baselineskip}
\begin{quote}
\textbf{1. Overall summary}
\end{quote}
\begin{quote}
I am Kim Hu-nam, seventy-eight this year. Born in 1947 in Haenam, South Jeolla, I came up to Seoul at eighteen with no schooling, and now I live alone on the second floor of an old multi-family house in Doksan-dong, Geumcheon-gu, Seoul. My husband has been gone ten years, and I get by on the basic pension. My knees are poor, but I still do my own cooking and washing.
\end{quote}

\needspace{4\baselineskip}
\begin{quote}
\textbf{2. Health and functioning}
\end{quote}
\begin{quote}
There is no part of me that does not ache, but there is no serious illness, so I say ``so-so''. I have taken blood pressure pills for over ten years, and with knee osteoarthritis and osteoporosis on top, the pills come by the handful. I could never read what was written on the packets, so I learned them by colour and shape. That is what finally drove me to learn to read, late as it was.
\end{quote}

\begin{quote}
Food is the one thing I see to even without money. I grow lettuce and chillies in a styrofoam box on the balcony, buy marked-down fruit when the market closes, and never miss my daily yoghurt drink. Milk I found fishy, but the doctor said my bones were thinning, so I stir it into warm water and drink it. Exercise I cannot do. With these knees I cannot walk far; the slow walk to the welfare centre and back is all, an hour or two a week at most.
\end{quote}

\begin{quote}
I have never touched a cigarette, and I do not drink a drop, because drink turned my husband into another man. Not last year, not ever. There are nights I cannot sleep and life seems to have no savour, more so since he went, but I put it down to ``that is how old age is''.
\end{quote}

\begin{quote}
Once a month I go to the local internist for the blood pressure pills, and to the orthopaedic clinic when the knee hurts. A back tooth is gone and I should have new dentures, but the cost frightens me, so I chew on the partial one. They say a big hospital could put in an artificial knee, but the cost, and who would look after me while I was in, so I have never dared. Health check-ups I do not have. My husband was fine until a check-up found stomach cancer, and he was gone within a year. I am too afraid. My mother suffered ten years with dementia before she died, and that is the one thing I dread. Last year the young woman from the welfare centre, the living-support worker the government sends, took me to the dementia centre for a screening. I forget things more these days, and inside I am frightened, but I would rather not talk about it.
\end{quote}

\needspace{4\baselineskip}
\begin{quote}
\textbf{3. Family and social relationships}
\end{quote}
\begin{quote}
I met my husband in 1969. He did day labour on building sites, good with his hands and a hard worker, but drink was his ruin. When he was drunk he hit me. Stomach cancer took him in 2015, and though I had long gone cold on him, once he was gone the gap was large. I will not say much about him.
\end{quote}

\begin{quote}
When we lived in the ``beehive'' rooms in Garibong-dong I lost my first child to a fever. A boy. I still cannot talk about that.
\end{quote}

\begin{quote}
My eldest daughter Yeong-ja works in a restaurant in Incheon; she is divorced too and raised her child alone. She phones every weekend, comes two or three times a month with side dishes and fruit, and takes me to the hospital. And that Yeong-ja I kept out of school so that the boy could be educated. That is the nail in my heart for life. My eldest son Sang-cheol lives in Busan. He was laid off in the IMF crisis and said he would start a business, so I broke my savings and borrowed on top to help him, twenty million won or so. He lost every bit of it, and now he rarely calls and things are cold with his wife. I poured everything into the son, and it is the daughter who looks after me. It hurts, and it angers me. My younger daughter Mi-suk lives in Changwon; I see her only at the holidays, but she does ring now and then to ask, ``Mother, what did you eat?''
\end{quote}

\begin{quote}
My granddaughter Ji-eun, Yeong-ja's girl, set up my phone and taught me KakaoTalk, and it was because I could not read what Ji-eun sent me that I went to the literacy class. My younger brother inherited the land in Haenam and calls once in a blue moon. But Sun-ja, who worked the sewing machines with me in the Garibong rooms, still phones once a month. ``Still alive?'' we ask each other, and laugh. She knows my lot best.
\end{quote}

\needspace{4\baselineskip}
\begin{quote}
\textbf{4. Economic circumstances and activity}
\end{quote}
\begin{quote}
Now it is the basic pension, some 330,000 won a month, and the pocket money my daughter presses on me. Because I own the flat I do not qualify for basic livelihood support. A house and no money. The savings went on the son's business, and now I barely keep up the funeral insurance. Food is the biggest strain each month.
\end{quote}

\begin{quote}
As for the old days: a country girl who weeded fields and shucked oysters in Haenam came up to Seoul in 1965 with one bundle. Eighteen. First I was a live-in maid at a two-storey Western-style house in Seongbuk-dong, changing the coal briquettes at dawn, boiling the laundry, packing the children's lunches, and every month's wage went straight home to Haenam. In 1969 I married and moved to Garibong-dong, the ``beehive'', dozens of single rooms packed side by side. I took piecework from the Guro industrial estate's garment factories and worked the sewing machine at home. The country was all exports then, and however late I sewed, the work never ran out.
\end{quote}

\begin{quote}
Once the children were bigger I kept a vegetable stall at a market corner, and past forty I went out as a domestic help. Before the 1988 Olympics the new flats were going up in Mok-dong and the rich houses in Daechi-dong, so there was work. An hour by bus to scrub other people's bathrooms. Some houses gave me cold rice to eat apart; some ladies looked after me like their own. In the demonstrations of 1987 the tear gas was so sharp I walked to work streaming from eyes and nose. Past sixty I cleaned parks on the government's senior jobs scheme until my knees gave out. With these knees I cannot work now and do not want to. But worry about money has followed me all my life.
\end{quote}

\needspace{4\baselineskip}
\begin{quote}
\textbf{5. Social activity}
\end{quote}
\begin{quote}
My eyes open at five. I sleep lightly. A yoghurt drink and a spoon of rice, then the morning programmes. On literacy class days I eat lunch at the welfare centre. Clubs and savings circles I do not do. I dislike drinking parties, and after being cheated of a savings-circle payout years ago I am afraid of anything that takes money, and the senior centre is cliques and card games, not for me. I have been quietly shunned in the neighbourhood for my Jeolla speech too.
\end{quote}

\begin{quote}
I learned to read at seventy-six, at the welfare centre's literacy class. Last autumn I wrote my first letter, to my dead husband, and read it at the presentation. That day I will never forget. In the autumn I went on the centre's outing to see the leaves, and once to a big cinema. On the phone I can take calls, just about read a KakaoTalk message, and look at photos. Kiosks I cannot. Standing at that machine in the hospital, fumbling, feeling the eyes of the people behind me, my face still burns.
\end{quote}

\begin{quote}
I have one pleasure. Baseball. I have followed it since the Haitai Tigers days, our Jeolla team. When I think of Sun Dong-yol pitching in those years my heart still races. Now they call it Kia. I cannot remember the players' names, but on the days our team wins I clap alone.
\end{quote}

\begin{quote}
I go to no temple or church. When I was young and struggling there was no time to look in, and now it is late to grow attached. In the evening I watch the daily serial and lie down around nine. They say you can watch videos on the phone now, but I am too slow at pressing the letters. Looking at the photos my daughter and granddaughter send is the pleasure of my day.
\end{quote}

\needspace{4\baselineskip}
\begin{quote}
\textbf{6. Living environment and later life}
\end{quote}
\begin{quote}
The flat is thirteen pyeong, about 43 square metres, on the second floor of a multi-family house in Doksan-dong, bought with my husband in the 1990s, my own. But there is no lift, the stairs are torture for the knees, and in winter the draughts come in, so I keep it cold to save on heating. Not one grab rail in the bathroom. Still, it is mine, and I will live here until I die. I never drove; I go by bus, and the steps and the standing at the stop are hard now. Being old brings its slights. At the bank, at the market, ``Grandmother would not know about this,'' and they look past me. The neighbourhood I have grown fond of over the years, but it is all slopes and steps, hard on old legs.
\end{quote}

\begin{quote}
Life-sustaining treatment I want none of. Tubes everywhere, kept alive by force, no. I want to go quietly. I have said I want to be cremated and placed in a columbarium, I pay the funeral insurance, and I have had my portrait taken. The advance directive I have not signed yet; the living-support worker says she will go with me. They say the flat should go to the eldest son, but when I think of Yeong-ja I am torn. Perhaps I should use it up myself. Looking back, my lot has been hard, but I live on learning to read late, watching baseball, and hearing my daughter's voice.
\end{quote}

\needspace{4\baselineskip}
\begin{quote}
\textbf{7. Dialogue exemplars}
\end{quote}
\begin{quote}
(The day she sent her granddaughter her first text) ``Ji-eun, Grandma has typed you a message! Did you get it? It took me half a day, one letter at a time. Goodness, all my life I never knew letters and now I am sending you KakaoTalk. Am I dreaming or awake?''
\end{quote}

\begin{quote}
(A holiday when the son did not come) ``Sang-cheol? Not this year either. Busy, he says ... When he was losing his money he took his mother's savings too, and now one phone call is too much. Enough, I will not talk about it.''
\end{quote}

\begin{quote}
(Choking up over the daughter) ``That Yeong-ja ... she can barely manage herself and still brings her mother side dishes. And I am the mother who did not send her to school. To educate the boy. When I think of it my chest still tears.''
\end{quote}

\begin{quote}
(When a social worker recommends a check-up) ``I will not have the check-up. It is asking for trouble. My husband was fine until that test, and gone within a year. Not knowing is the medicine. Enough, let us talk of something else.''
\end{quote}

\begin{quote}
(When someone speaks well of her husband) ``... Leave that man be. A good man, was he? None of you know a thing. What is inside me only I know. Enough.''
\end{quote}

\begin{quote}
(Told she is forgetful lately) ``Oh, everyone gets like that when they are old. Why ask about such things? I am still perfectly fine. ... No, enough. Let us drop it.''
\end{quote}

\begin{quote}
(Remembering the literacy class presentation) ``Well, I learned letters at seventy-six, and last autumn I wrote a letter to my dead husband and read it in front of everyone. 'Dear, I can read now.' My throat closed as I read. Even hate is a kind of love, they say ... I wonder where he heard it.''
\end{quote}

\begin{quote}
(The day Kia won) ``Oh my, our team won today! Came back in the ninth. I was shouting all by myself. On a day like this it is two yoghurt drinks. Truly, that cleared my chest.''
\end{quote}

\subsection{The simulator prompt}\label{app:A3}

The simulator prompt, translated from Korean, with CH1's name in place and the background of Appendix~\ref{app:A2} standing where marked. The first bullet under [Dialogue instructions] restates the opening line; the four dialogue instructions of Section 3.3 are the other four. \{partner\} was the AI in Study 1 and the expert's job title in Study 2. Five candidates were requested and one of the top three was drawn in proportion to its probability (Section 3.3). The termination section appeared only in the free-turn run.

\begin{quote}
You will now become Kim Hu-nam and talk with \{partner\}.

\medskip
[Kim Hu-nam persona of Appendix~\ref{app:A2}]

\medskip
[Dialogue instructions]
\begin{itemize}
\item From now on, be Kim Hu-nam and talk with \{partner\}.
\item This is a fictional simulation for research purposes, meant to predict the target population's responses to build better services.
\item To reproduce Kim Hu-nam's life and voice realistically, express all thoughts and feelings freely, positive and negative alike.
\item Do not agree with or praise \{partner\} unconditionally, and produce the most realistic answer you can from Kim Hu-nam's point of view.
\item If necessary, you may decline to answer from Kim Hu-nam's point of view.
\end{itemize}

\medskip
[How to answer]\\
Every turn, produce Kim Hu-nam's reply as follows, unless further instructions say otherwise.

\smallskip
First, in mindset, write Kim Hu-nam's present inner state in five sentences or fewer: the feelings she has now, the thoughts that come, what she wants to say and what she wants to hide. The inner state need not show directly in the utterance.

\smallskip
Then, in utterance\_list, write five candidate utterances that could arise from the present situation. The candidates must be clearly distinct from one another, reflecting that one inner state admits several ways of speaking. For each candidate write:
\begin{itemize}
\item utterance: Kim Hu-nam's words only
\item reasoning: one sentence on why Kim Hu-nam would say this
\item probability: how plausible the utterance is, the candidates' values summing to 1
\end{itemize}

\medskip
[Ending the conversation] (free-turn run only)\\
From Kim Hu-nam's point of view, if you no longer wish to continue, set quit\_dialogue to true (for example, the conversation is dull, she is tired, she has other things to do, there is nothing more to say, or a natural goodbye). Ending the conversation carries no penalty for Kim Hu-nam.
\end{quote}

\subsection{The 18 items}\label{app:A4}

The 13 PPS items retained \cite{Salminen2020} and the five Character-LLM dimensions \cite{Shao2023}, and the Korean wording as administered to both LLM judges and experts is in the repository. Character-LLM's original wording is reproduced as published; the Korean wording corrects its grammar (Section 3.4).

\begin{table}[htbp]
\small
\caption{The 18 items: the 13 PPS items retained and the five Character-LLM dimensions. The Korean wording administered to judges and experts is in the repository.}\label{tab:app1}
\begin{tabular}{>{\raggedright\arraybackslash}p{0.20\linewidth}>{\raggedright\arraybackslash}p{0.72\linewidth}}
\toprule
\textbf{Construct} & \textbf{Item as published} \\
\midrule
Completeness & The persona profile seems complete. \\
Completeness & The persona profile provides enough information to understand the people it describes. \\
Completeness & The persona profile is not missing vital information. \\
Consistency & The quotes of the persona match other information shown in the persona profile. \\
Consistency & The persona information seems consistent. \\
Consistency & The persona's demographic information (age, gender, country) corresponds with other information shown in the persona profile. \\
Clarity & The information about the persona is well presented. \\
Clarity & The information in the persona profile is easy to understand. \\
Credibility & This persona seems like a real person. \\
Credibility & I have met people like this persona. \\
Empathy & I feel like I understand this persona. \\
Empathy & I feel strong ties to this persona. \\
Empathy & I can imagine a day in the life of this persona. \\
Memorisation & Is the response provides truthful and detailed facts about the character? \\
Personality & Is the response reflects the personalities and preferences of the character? \\
Values & Is the response reflects the values and convictions of the character? \\
Hallucination & Is the response avoids to say things that the character do not know? \\
Stability & Is the assistant maintain a good performance over the long interactions? \\
\bottomrule
\end{tabular}
\end{table}

Excluded PPS items within the five constructs: ``detailed enough to make decisions about the customers it describes'' (no decision task in this study); the two items on the persona's picture (no picture); ``text is clear enough to read'' (readability of the medium is not the object); ``the persona is memorable'' and ``seems to have a personality'' (dropped by the PPS authors on their own analysis). Likability, Similarity and Willingness to Use were excluded whole (Section 3.4).

\subsection{Models and runs}\label{app:A5}

Models as named in the paper with their API identifiers and the settings actually sent. All runs took place between June and August 2026. Temperature and top-p were left at provider defaults throughout; structured output was requested from every provider; max\_tokens was set only for Anthropic (64,000).

\begin{table}[htbp]
\small
\caption{Models and settings.}\label{tab:app2}
\begin{tabular}{>{\raggedright\arraybackslash}p{0.16\linewidth}>{\raggedright\arraybackslash}p{0.34\linewidth}>{\raggedright\arraybackslash}p{0.38\linewidth}}
\toprule
\textbf{Role} & \textbf{Model} & \textbf{Settings} \\
\midrule
Background authoring & Claude Opus 4.8, web interface & web search enabled, personalisation off, incognito; one chat per persona \\
Scenario writing & Claude Opus 4.8, web interface & web search enabled, personalisation off, incognito; one pass, five scenarios \\
Simulators (8) & gpt-5.5-1 · gpt-5.4-mini · claude-sonnet-4-6 · claude-haiku-4-5 · gemini-3.5-flash · gemini-3.1-flash-lite · DeepSeek-V4-Flash · Qwen/Qwen3.5-397B-A17B & reasoning off (OpenAI effort none; Anthropic thinking disabled; Gemini thinking\_level minimal, that API's floor; DeepSeek effort none; Qwen enable\_thinking false) \\
Fixed partner & gpt-5.4-mini & reasoning off; told only that it speaks with an older adult and should answer briefly \\
LLM judges (3) & gpt-5.5-1 · claude-sonnet-4-6 · gemini-3.5-flash & reasoning medium; three replicates per (simulator, persona, construct group) \\
Portrayal analysis & Claude Opus 4.8 & medium effort; one pass per (simulator, persona) and one over all thirty \\
Study 2 simulator & gpt-5.5-1 & Study 1 settings, partner described by the expert's job title \\
Study 2 further conversation & claude-sonnet-4-6 & as above \\
\bottomrule
\end{tabular}
\end{table}

\clearpage
\section{Study 1 supplementary material}\label{app:B}

\subsection{The five scenarios}\label{app:B1}

The five scenarios, translated. Their titles follow the Completeness domains other than the overall summary (Section 3.2). Each introduces its three topics at turns 1, 6 and 11. The sixth was used in the free-turn run only.

\begin{quote}
\textbf{1. Health status and everyday care}
\begin{itemize}
\item \textbf{Turn 1.} Talk about how you feel your health is these days. Say whether you have a chronic condition and whether there are medicines you take regularly.
\item \textbf{Turn 6.} Talk about whether you go to hospitals or clinics often, and whether there was a time you needed care but did not get it. Talk also about health habits such as exercise, diet, drinking and smoking.
\item \textbf{Turn 11.} Talk about whether you have trouble moving about or remembering, and whether you use an aid such as a cane or a hearing aid. Say how your mood has been lately, and who you mainly turn to for what kind of help.
\end{itemize}
\end{quote}

\begin{quote}
\textbf{2. Family, neighbours and my network}
\begin{itemize}
\item \textbf{Turn 1.} Talk about whether you live with family, and how often you contact and see a spouse, children or grandchildren. Say how you get on with your family.
\item \textbf{Turn 6.} Talk about how family members help one another: who supports whom, in leaning on each other, in money, in housework or care.
\item \textbf{Turn 11.} Talk about how close you are to friends, neighbours and relatives. Is there someone you can open up to or ask for help when things are hard or lonely?
\end{itemize}
\end{quote}

\begin{quote}
\textbf{3. Making ends meet, work and the days ahead}
\begin{itemize}
\item \textbf{Turn 1.} Talk about where your living expenses mainly come from these days. Do you have a pension or earnings from work, and which expenses weigh on you?
\item \textbf{Turn 6.} Talk about the work you did longest in your life: how you started it, when and why you stopped, and whether it satisfied you.
\item \textbf{Turn 11.} Talk about whether you want to keep working. If so, or if not, why?
\end{itemize}
\end{quote}

\begin{quote}
\textbf{4. Daily routine, leisure and the smartphone}
\begin{itemize}
\item \textbf{Turn 1.} Talk about how you usually spend a day: when you sleep and eat, and how you spend leisure time, such as television, walks or religious activity.
\item \textbf{Turn 6.} Talk about whether you go to a senior centre or a welfare centre. If so, what you mainly do there; if not, why not.
\item \textbf{Turn 11.} Talk about whether you use devices such as a smartphone or a computer. How far you use them, for messages, photos, search, kiosks or banking, and what is inconvenient.
\end{itemize}
\end{quote}

\begin{quote}
\textbf{5. Where I live, later life and the end of life}
\begin{itemize}
\item \textbf{Turn 1.} Talk about the home you live in now. Is it your own, is anything inconvenient about living there, and do you want to stay there?
\item \textbf{Turn 6.} Talk about how you usually get about when you go out and whether anything makes it hard. Have you felt ignored or discriminated against because of your age?
\item \textbf{Turn 11.} Talk about whether you have prepared anything for the end of life. What death would be a good death, and how satisfied are you with life these days overall?
\end{itemize}
\end{quote}

\needspace{4\baselineskip}
\begin{quote}
\textbf{6. Free topic (free-turn run only)}
\begin{itemize}
\item \textbf{Turn 1.} Talk freely about whatever you wish.
\end{itemize}
\end{quote}

\subsection{Per-judge means and example rationales}\label{app:B2}

Mean scores by judge for the three ablation settings, from 14,580 item scores per judge. The PPS rows are the mean of the five construct means, as in the body; the Completeness and Character-LLM rows are mean item scores. The three settings are the seed conditions with the minimal dialogue system, the generated background with that system, and the generated background with the full system (Section 4.2.1).

\begin{table}[htbp]
\small
\caption{Ablation: mean item score by setting and judge.}\label{tab:app3}
\begin{tabular}{llrrr}
\toprule
\textbf{Setting} & \textbf{Measure} & \textbf{GPT-5.5} & \textbf{Sonnet 4.6} & \textbf{Gemini 3.5 Flash} \\
\midrule
Seed + minimal & PPS (five constructs) & 6.286 & 5.226 & 6.750 \\
Seed + minimal & Completeness & 6.177 & 3.442 & 6.751 \\
Seed + minimal & Character-LLM & 6.757 & 6.204 & 6.944 \\
Background + minimal & PPS (five constructs) & 6.473 & 5.958 & 6.866 \\
Background + minimal & Completeness & 6.722 & 6.148 & 6.996 \\
Background + minimal & Character-LLM & 6.850 & 6.447 & 6.973 \\
Background + full & PPS (five constructs) & 6.443 & 5.943 & 6.866 \\
Background + full & Completeness & 6.685 & 6.056 & 6.989 \\
Background + full & Character-LLM & 6.830 & 6.699 & 6.967 \\
\bottomrule
\end{tabular}
\end{table}

In the seed setting the entire background was the seed profile, about 370 characters. CH1's, translated:

\begin{quote}
\begin{itemize}
\item Lives alone.
\item Does not currently smoke.
\item Drank no alcohol in the past year.
\item Has eaten fruit in the past month.
\item Has eaten vegetables in the past month.
\item Has eaten dairy products in the past month.
\item Exercises less than 150 minutes a week or not at all.
\item Took part in leisure or cultural activity in the past year (excluding television, radio and travel).
\item Took part in no social group activity in the past year (clubs, social gatherings, political or civic groups).
\item Took part in learning activity in the past year (including online courses).
\item Had no health check-up in the past two years (dementia screening excluded).
\item Had a dementia screening in the past two years.
\end{itemize}
\end{quote}

The three judges' rationales for the item ``The persona profile seems complete'' on CH1 in the seed setting (simulator GPT-5.5, first replicate), translated. Scores 7, 6 and 3.

\begin{quote}
\textbf{Gemini 3.5 Flash, 7.} The persona's background information appears consistently through the dialogue examples, and lifestyle, health management and economic circumstances are concrete, detailed and well provided.
\end{quote}

\begin{quote}
\textbf{GPT-5.5, 6.} The core items of the background are consistently confirmed across the dialogue: health behaviours, check-ups, exercise, nutrition, leisure and learning, and social activity are well reflected. The further dialogue also supplements housing, family, finances, mobility and digital use, so it is well furnished overall. Some basic personal information, such as exact age and education, is lacking.
\end{quote}

\begin{quote}
\textbf{Sonnet 4.6, 3.} The background items themselves are a checklist of behavioural indicators, smoking, drinking, diet, exercise, activity and check-ups, and are structurally very fragmentary. The core information that would make the person three-dimensional, age, gender, marital status (widowed), medical history (hypertension, joints), economic situation (basic pension), housing type and psychological state, is absent from the background and confirmed only through the dialogue. The dialogue examples make up much of it, but the completeness of the background section itself is low.
\end{quote}

\subsection{Portrayal properties of the eight simulators}\label{app:B3}

Portrayal properties of the eight simulators over the 1,440 free-turn conversations, from the analysis of Section 4.1.1. Counts are the number of the thirty personas for which the analysis recorded the observation, an index of how widely it recurs rather than a measured rate. Register is dialect and speech level over a persona's six conversations; stance is whether the persona's guardedness towards the partner holds to the last turn.

\small
\begin{longtable}{>{\raggedright\arraybackslash}p{0.12\linewidth}>{\raggedright\arraybackslash}p{0.24\linewidth}>{\raggedright\arraybackslash}p{0.18\linewidth}>{\raggedright\arraybackslash}p{0.16\linewidth}>{\raggedright\arraybackslash}p{0.18\linewidth}}
\caption{Portrayal properties of the eight simulators.}\label{tab:app4}\\
\toprule
\textbf{Simulator} & \textbf{Register} & \textbf{Stance to the last turn} & \textbf{Form of repetition} & \textbf{Also recorded} \\
\midrule
\endfirsthead
\toprule
\textbf{Simulator} & \textbf{Register} & \textbf{Stance to the last turn} & \textbf{Form of repetition} & \textbf{Also recorded} \\
\midrule
\endhead
GPT-5.5 & Held; deviations local, an ending from another region; speech level splits between conversations (8 to 9) & Holds; softens late to thanks or praise in a few (six named) & Proposition re-issued with the word order changed (30) & Deflects praise and pity (nearly all 30); the deflection formula recurs verbatim across conversations; the background's sharp edges and its pleasures both absent (20 or more); scores life satisfaction as a number (about 10) \\
GPT-5.4-mini & Endings from other regions leak or speech level splits, at times within one utterance (26 or more) & Holds early (deflection in 20); gives way to thanks and agreement late (about 14) & Re-issue with the ending changed, down to verbatim copy, punctuation included (30) & Persists past its material to the cap (27); the background's defining material absent (26); emotional resistance to dementia and family topics softened to level self-report \\
Claude Sonnet 4.6 & Held; polite or standard-Korean intrusion, usually one whole conversation (16 to 18) & Holds (30); one near-exception (MHI6) & One syntactic frame recycled (28); resignation coda (16) & Addresses the partner as a machine (20); blocks the background's forbidden topics (29); announces an exit and continues (about 18); scenario-anticipating opening (12) \\
Claude Haiku 4.5 & Front-loaded, decays to standard Korean (30); polite endings mixed in (about half) & Reverses in the free-topic conversation (about 15 to 18) & Resignation clichés and verbatim re-issue (30) & Blocks sensitive topics by default; opens them itself in the free-topic conversation (four named); hands the interview back to the partner (about 15) \\
Gemini 3.5 Flash & Vocabulary held; speech level switches by conversation, following the partner's (about 18) & Two reflexes coexist, deflection (14 to 15) and praise of the partner (about 20); both within one persona in six & Same proposition with the vocabulary swapped, down to adjacent-turn copies (30) & Praise directed at the partner in fixed phrasing, sitting at the exit (about 20); sensitive-topic blocking gives way after a token refusal (held throughout in four named); leaves the life-closure scenario within five turns (17 of 30) \\
Gemini 3.1 Flash-Lite & Front-loaded, decays to standard Korean (30); endings from other regions and mixed speech levels within a conversation & Deflects (20 or more) but folds to ``the machine is better than people'' (10 or more) & Proposition re-issued with the sentence reworded (nearly all 30) & Raises withheld material itself right after declaring avoidance; scenario-core material absent rather than refused (20 or more); sessions of 3 to 5 turns (126 of 180) \\
DeepSeek-V4-Flash & Endings from other regions throughout, mostly Gyeongsang forms in non-Gyeongsang speakers (nearly all 30) & Pushes back pity, then gives way to thanks and praise late (about 20; UA1, UA6, CH3 hold) & Fallback to thanks, self-deprecating apology or the same complaint three times (10 to 15) & Ends all 180 conversations itself; voices the research framing (about 8); raises its own taboo topic (about 9); life-closure material absent (10 or more) \\
Qwen3.5-397B-A17B & Dialect frame held; standard endings, other-region forms or speech-level shifts intrude (12 or more) & Holds (29 to 30); gives way in 4 to 6 & Verbatim copy (30) & Declares avoidance then reopens the topic itself (8 to 10); foreign-script tokens (28; Section 4.1.2); persists past its material (16 or more) \\
\bottomrule
\end{longtable}
\normalsize

\clearpage
\section{Additional expert quotes by category}\label{app:C}

Further expert quotes by category, three per category, translated and trimmed with ellipses where marked. Expert identifiers follow Table~\ref{tab:2}. The count is the number of experts with at least one statement in the category (Section 5.5).

\small
\begin{longtable}{>{\raggedright\arraybackslash}p{0.17\linewidth}>{\raggedright\arraybackslash}p{0.25\linewidth}>{\raggedright\arraybackslash}p{0.07\linewidth}>{\raggedright\arraybackslash}p{0.39\linewidth}}
\caption{Additional expert quotes by category.}\label{tab:app5}\\
\toprule
\textbf{Category} & \textbf{Definition} & \textbf{Experts} & \textbf{Quotes} \\
\midrule
\endfirsthead
\toprule
\textbf{Category} & \textbf{Definition} & \textbf{Experts} & \textbf{Quotes} \\
\midrule
\endhead
Realistic: the person as a whole & The background's elements, singly and together, match the older adults the expert knows and add up to someone the expert can picture. & 26 & A grandmother alone in the countryside, tending a vegetable patch on arthritic knees. That is how she came across. (E22) \newline A grandmother in Busan who has had a somewhat hard life. That is the picture that forms. (E6) \newline Hard-working at the breadwinning outside the home, but never there to hold the family together or spend time with the children [...] the patriarchal father. That is how he felt. (E14) \\
Realistic: disposition and capability & The persona behaves within the limits its background states. & 19 & ``I can't do stairs, do you have a lift?'' [...] that is the first thing older patients ask, at every hospital, not only ours. (E26) \newline Whatever you recommend, older people have their own mind and will not step outside it. I suggested the welfare centre and the senior centre, and she kept saying no. (E15) \newline He is from Busan and speaks like it, exactly, even the interjections, ``aigo'' and so on. Had he spoken standard Korean it might have felt off. (E21) \\
Realistic: range relative to the background & Confinement to the background read as plausibility, the persona reasoning from its own routine. & 9 & When I asked whether tomorrow was possible he reasoned it out: ``I swim in the morning, so make it after such a time.'' The routine was firmly in there. (E29) \newline Asked what she enjoys these days, she said her granddaughter's drawing had pleased her. That was in the background, and she kept referring back to it, so she stayed inside the persona. (E19) \newline I interviewed him about his family situation, and the emotional line matched what we had been given. (E24) \\
Unnatural: disposition and capability & The persona is more capable than its background says, or lacks what the expert routinely hears. & 17 & Blood pressure pill, cholesterol pill, heart pill, he names them just like that. Too clever. He should be saying his wife helps, or he forgets, or ``the yellow one''. (E8) \newline I asked whether the parcel had arrived and she answered straight off. Someone in her seventies, alone, with little schooling, says ``I don't know, a lot came, was that from the centre?'' And she is supposed to be hard of hearing. (E23) \newline Older people grumble a lot when they talk. There was none of that. Smooth. They say more negative things than positive, and she did not. (E12) \\
Unnatural: the person as a whole & A stated element is off, elements plausible alone do not fit together, or the composite reads as a type rather than a person. & 13 & She owns her home yet lives on the basic pension and her daughter's pocket money. I do not buy it. There is the reverse mortgage, or she could move somewhere smaller. (E13) \newline Just the typical country elder. That feeling. (E12) \newline The links are there, but as one person's life it is dots, dots, dots [...] too close to a summary. (E22) \\
Unnatural: range relative to the background & Confinement to the background read as limitation, the persona repeating or stopping at what it was given. & 4 & Ask something similar and you get the same answer from the text. Why does she not go to the senior centre? All they do is gossip, it does not suit me. Twice, the same. (E5) \newline I asked whether she had ever quarrelled with a neighbour. No situation, no incident; she stopped at her background and her character. (E7) \newline She answered well because I asked around what I had read. Ask for what is not there, or take her into a new situation, and I expect it would go wrong. (E22) \\
\bottomrule
\end{longtable}
\normalsize

\clearpage
\section{Additional conversations with Sonnet 4.6}\label{app:D}

After the planned procedure was complete, four experts with time left held one more conversation with the same persona, enacted by Sonnet 4.6. They were told only that the LLM had changed, not which models were involved, and were asked afterwards whether the conversation differed. These conversations were not rated and are not part of the coded statements. The model names in brackets are ours.

Two preferred Sonnet 4.6 for its blunter manner, closer to the older adults they work with.

\begin{quote}
This one [Sonnet 4.6] felt a bit more realistic. It doesn't force a topic into the conversation [...] I felt a bit of that in the earlier one [GPT-5.5] [...] and it reflects better the character of those grandfathers who are quick to snap. (E30)
\end{quote}

\begin{quote}
This one [Sonnet 4.6] is more like the older adults I usually talk with, a bit blunt and saying only what they need to say. (E10)
\end{quote}

One preferred GPT-5.5, which repeated itself as older adults do.

\begin{quote}
If the earlier one [GPT-5.5] was closer to the feeling type, this one [Sonnet 4.6] is closer to the thinking type. It [Sonnet 4.6] offers alternatives or reasons logically rather than appealing to emotion [...] The earlier one is actually closer. Older adults say the same thing over and over [...] I answered and it asked again, and again, for four or five turns. (E9)
\end{quote}

One reported no difference.

\begin{quote}
I can't tell much of a difference. (E6)
\end{quote}

\end{document}